\newcommand{\siamls}[1]{#1} 
\renewcommand{\siamls}[1]{} 

\documentclass[11pt]{article}

\siamls{%
  \usepackage[mathlines]{lineno}
  \linenumbers
}

\usepackage{times}
\usepackage[version=4]{mhchem}
\usepackage[numbers,comma]{natbib}

\usepackage{caption}
\usepackage[numbers,comma]{natbib}

\DeclareMathOperator{\supp}{supp}

\usepackage{amsmath,amssymb}
\usepackage{array}
\usepackage{booktabs}
\usepackage{mathtools}
\usepackage{amsthm}  
\newcommand{\FIGDIR}{FIGURES}
\newcommand{\pic}[2]{\includegraphics[scale=#1]{\FIGDIR/#2}}
\newcommand{\picc}[2]{\begin{center}\pic{#1}{#2}\end{center}}
\newcommand{\minp}[2]{\begin{minipage}{#1\textwidth}#2\end{minipage}}

\newcommand{\comment}[1]{}
\newcommand{\edo}{\end{document}}

\usepackage[margin=1in]{geometry}
\theoremstyle{plain}
\newtheorem{theorem}{Theorem}
\newtheorem{lemma}{Lemma}

\theoremstyle{remark}
\newtheorem{remark}{Remark}

\newtheorem*{proof*}{Proof}

\newcommand{\SY}{Y}
\newcommand{\SA}{A}
\newcommand{\SB}{B}
\newcommand{\SC}{C}

\newcommand{\SYABC}{YABC}

\newcommand{\be}{\begin{equation}}
\newcommand{\ee}{\end{equation}}
\newcommand{\beqn}{\begin{eqnarray*}}
\newcommand{\eeqn}{\end{eqnarray*}}
\newcommand{\bi}{\begin{itemize}}
\newcommand{\ei}{\end{itemize}}
\newcommand{\ben}{\begin{enumerate}}
\newcommand{\een}{\end{enumerate}}

\newcommand{\bR}{\mathbf{R}}
\newcommand{\maybestar}{} 

\title{Dynamics and dose response in scaffold ligand binding}
\author{Eduardo D. Sontag \\
  Northeastern University\\
Departments of Electrical and Computer Engineering and
BioEngineering\\
Affiliate, Departments of Mathematics and Chemical Engineering}
\date{}

\begin{document}
\maketitle

\begin{abstract}
  \noindent
This paper considers systems in which two or more ligands bind
independently to distinct sites in a common scaffold. Such systems
arise in a range of applications, including immunotherapy and
synthetic biology. We show that each stoichiometric compatibility
class contains a unique steady state, and that this steady state is
asymptotically stable. The main result gives a rigorous proof that the
steady-state concentration of the fully bound complex, viewed as a
function of the total scaffold concentration, has a unique maximum. This biphasic dose response behavior is a characteristic feature of scaffolding systems and, in the special case of two ligands, plays an important role in the design and analysis of bispecific antibody drugs.

\siamls{%
  \noindent\textit{Relevance to Life Sciences:}
From bispecific antibodies for cancer therapy, to CRISPRa gene circuits, to cell signaling and enzymology, scaffold-mediated binding plays a central role in immunology, intracellular signaling, and synthetic biology. This paper shows that, for any number of ligands ($m\ge 2$), the steady-state concentration of the fully bound scaffold--ligand complex has a unique maximum at an intermediate scaffold concentration. This prediction is directly relevant to the emerging modality of trispecific antibodies, now being developed by several major pharmaceutical companies. We also prove results on the monotone, biphasic, and in some cases multiphasic behavior of partially bound complexes. These predictions should be experimentally testable and, when multiple local maxima occur, may have implications for dose selection and therapeutic optimization.

\noindent\textit{Mathematical Content:} The dynamics of scaffold systems are
analyzed using tools from reaction network theory, Petri nets, and
Feinberg theory. The study of biphasic behavior employs just
elementary calculus concepts and algebra, yet the proof is quite
complicated and it required a number of variable transforms.
    
}

\end{abstract}  


\section{Introduction}

\begin{figure}[ht]
\picc{0.5}{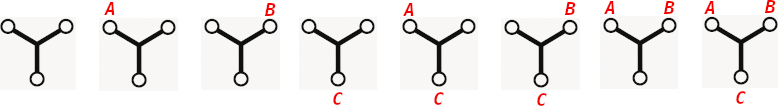}
\caption{Scaffold with three types of ligands.
  Here ligands are denoted by $\SA,\SB,\SC$ and the scaffold is
  denoted by $\SY$. 
  There are eight possible configurations, ranging from the free
  scaffold to the fully loaded one.
}
\label{fig:generic_scaffold}
\end{figure}

This paper studies a dynamical system that describes the binding and
unbinding of $m$ types of ligands to a connecting scaffold.
These ligands can independently bind to $m$ specific sites on the
scaffold.
The possible configurations of scaffold and ligands are shown, for the
special case $m=3$, in Figure~\ref{fig:generic_scaffold}.

Our first result shows that all positive solutions converge to a uniquely defined steady state within their stoichiometric compatibility class.
The total concentrations of scaffold and ligands (in any of their bound forms)
are conserved along solutions, and equal the initial concentrations of the
respective unbound species.
For any such total numbers, one has what is called a stoichiometric conservation class.
Theorem~\ref{theo:global_general}, given in Section~\ref{sec:dynamics},
proves there is a unique steady
state in each conservation class, and it is globally
asymptotically stable relative to the class. The proof relies upon
showing ``detailed balance'' of the chemical reaction network (CRN)
associated to these reactions, together with basic theory of
interaction networks including the theory of siphons and persistence.

In many applications, it is important to understand what the
``optimal'' scaffold dose $Y_{\text{tot}}$ is, in the sense of that maximizing
the steady state concentration of the fully formed complex (for
example, $\SYABC$ when $m=3$).
The intuition is that for small amounts of $Y_{\text{tot}}$ there will
be few complexes possible, but, for very large amounts of scaffold
compared to ligands, ligands will rapidly bind to separate scaffolds,
thus not allowing the opportunity for complete complexes to form.
One might view this biphasic behavior as reflecting a competition among
the ligands for binding to a ``resource'' (scaffold).
In Section~\ref{sec:biphasic}, we provide our main result,
Theorem~\ref{theo:biphasic_general}, which shows that this biphasic behavior is
always true, as long as there are $m\geq2$ ligands.

In Section~\ref{sec:monotonefree}, we complement
Theorem~\ref{theo:biphasic_general} with a third result,
Theorem~\ref{theo:monotonefree}, which establishes that each free
species depends monotonically on the total scaffold. The bound
complexes behave more richly: the fully loaded complex is biphasic,
and, as also discussed in Section~\ref{sec:monotonefree}, certain
partially loaded complexes depend on the total scaffold in an even
more intricate way, exhibiting two or more local maxima.

The rest of this paper is organized as follows.
Section~\ref{sec:motivation} explains the motivation for this study,
from both pharmacology (namely bi- and trispecific antibodies) and
synthetic biology (CRISPR activation constructs).
Section~\ref{sec:preliminaries} sets up basic terminology and
definitions, and Sections~\ref{sec:dynamics} to~\ref{sec:monotonefree}
present the main results.
We close in Section~\ref{sec:remarks} with some general remarks.

\section{Motivation}
\label{sec:motivation}

\begin{figure}[ht]
\minp{0.2}{\picc{0.15}{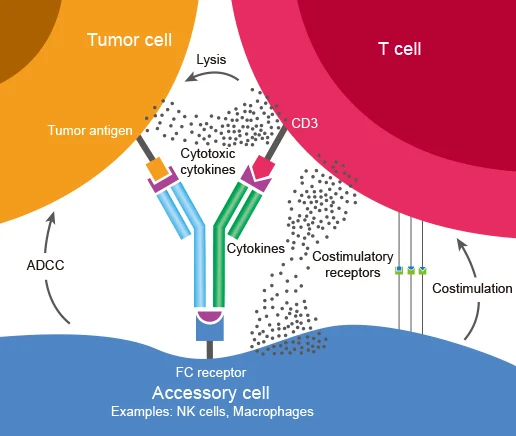}}%
\minp{0.4}{\picc{0.15}{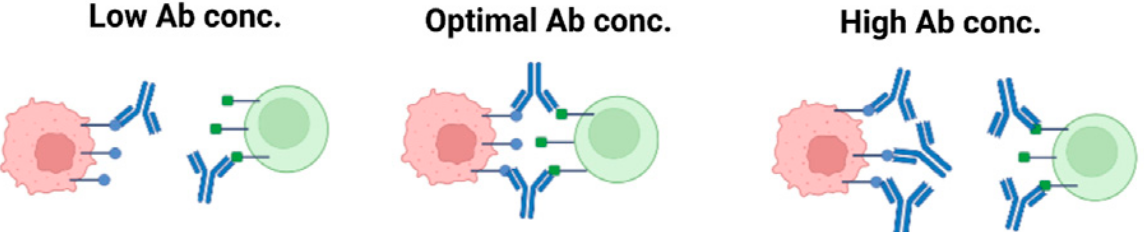}}%
\minp{0.25}{\picc{0.1}{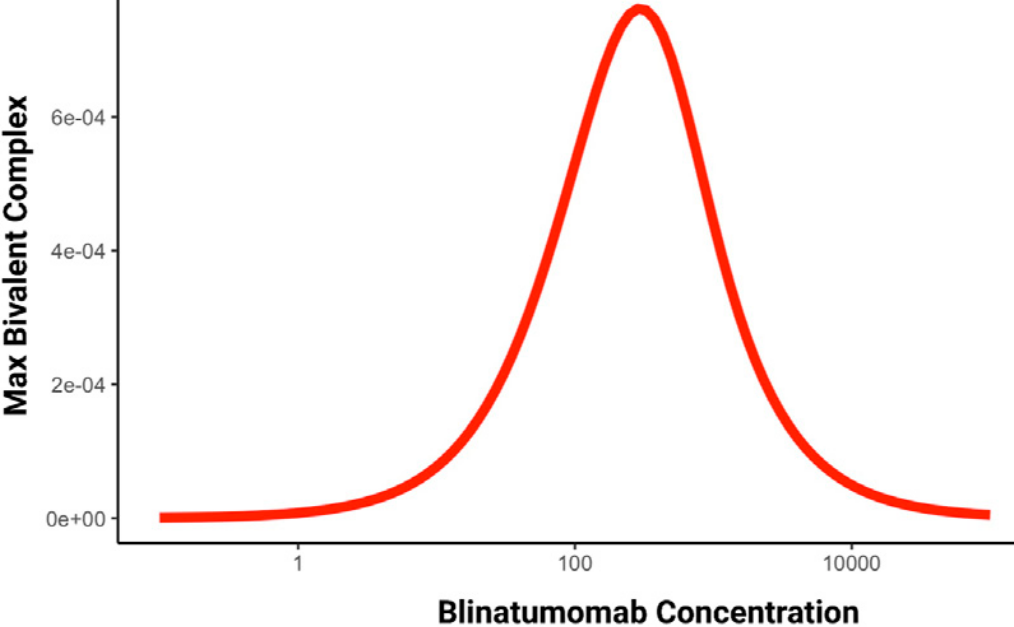}}%
\minp{0.1}{\picc{0.15}{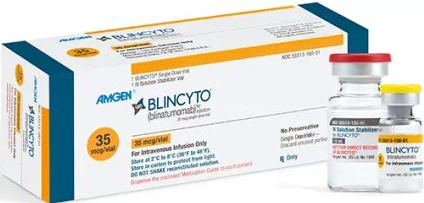}}
\caption{Bispecific T-cell Engagers (BiTE) constitute a promising class of drugs in oncology.
Left to right:
These antibodies have one Fab (Fragment antigen-binding) arm that
typically binds CD3 receptors in T cells, and another one that
binds a tumor-associated antigen such as HER2 or CD19 on a cancer
cell; Figure from~\cite{genscript}.
(The diagram shows also an Fc or Fragment crystallizable region that can
provide effector functions such as antibody-dependent cell-mediated
cytotoxicity but is not involved in antigen recognition or binding.)
There is an optimal concentration of antibody if the objective is to
maximize complex availability, as the diagram suggests.
An example is Blinatumomab, a targeted immunotherapy for B-cell acute
lymphoblastic leukemia (ALL); figure from~\cite{dunlapcao}.
Shown is the biphasic response; figure also from~\cite{dunlapcao}.
}
\label{fig:bispecific}
\end{figure}

Our motivation for studying this class of systems arose from two different applications.
The first one is the design of therapeutic bispecific antibodies.
See Figure~\ref{fig:bispecific} for an illustration.

\begin{figure}[ht]
\picc{0.20}{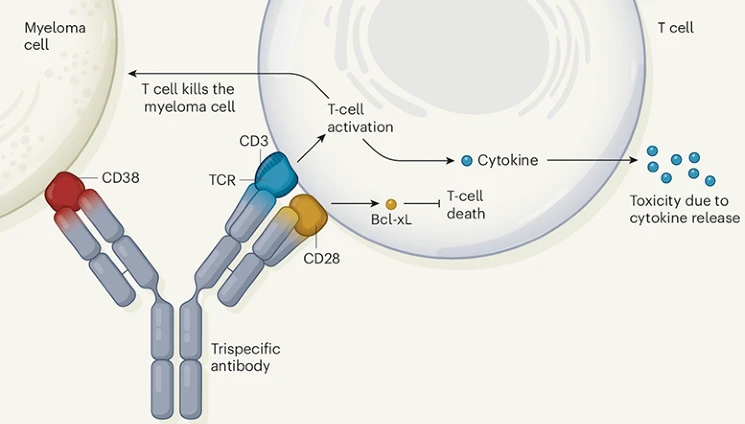}
\caption{Example of trispecific antibody, figure reproduced from~\cite{Wu2020_Trispecific_NatCancer}.}
\label{fig:trispecific}
\end{figure}

In this application, there are $m=2$ ligands, which are receptors in
cancer and immune cells, and the scaffold is the bispecific
antibody. For $m=2$, it was known 
(though complete proofs are not readily available) that the steady state
response is biphasic, reaching a maximum at exactly one value of 
total scaffold concentration.
See~\cite{2024_bispecific} for a recent paper where we studied the
dynamic behavior and identifiability of bispecific antibody interactions.
An even newer related therapy is that of trispecific antibodies, see
Figure~\ref{fig:trispecific}. In this case, $m=3$, and there may be
two receptors to bind in an immune or cancer cell, guaranteeing more
specificity and/or stronger effects.

\begin{figure}[ht]
\picc{0.75}{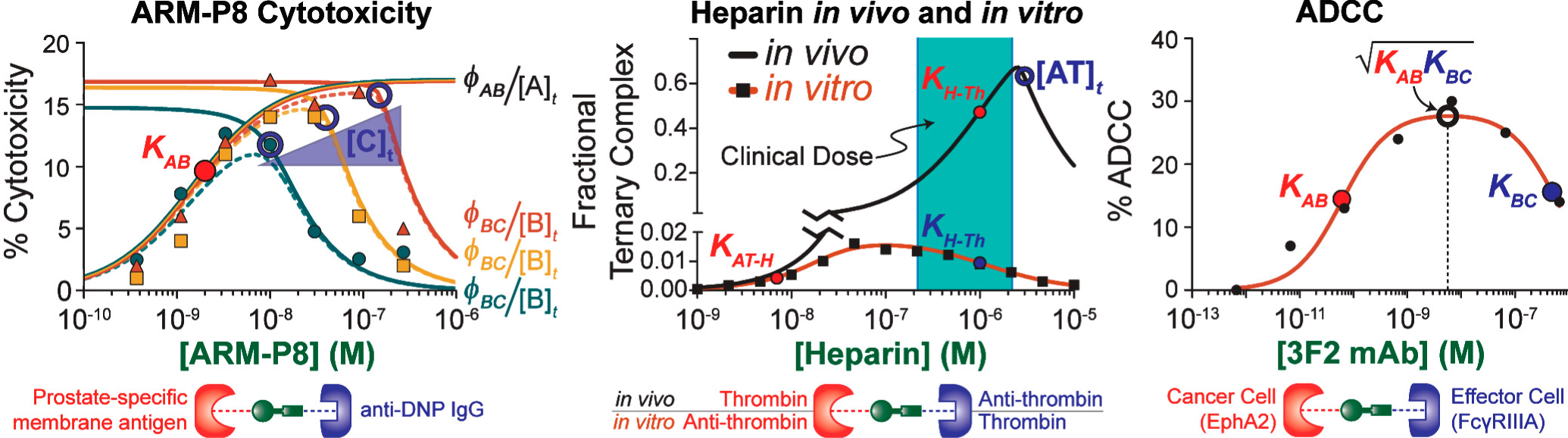}
\caption{Examples of biphasic responses due to scaffolding.
Left to right:
Antibody-recruiting small molecules (ARMs) that target prostate cancer:~PSA/IgG.
Heparin as bridge bringing antithrombin and thrombin together in a
ternary complex (antithrombin inactivates clotting factors such as thrombin).
A bispecific antibody.
Illustrations adapted from~\cite{Douglass2013ThreeBody}.}
\label{fig:other_biphasic}
\end{figure}

Other examples of biphasic responses are shown in
Figure~\ref{fig:other_biphasic}; see the discussion
in~\cite{Yang2011ScaffoldmediatedNO}.

\begin{figure}[ht]
\picc{0.5}{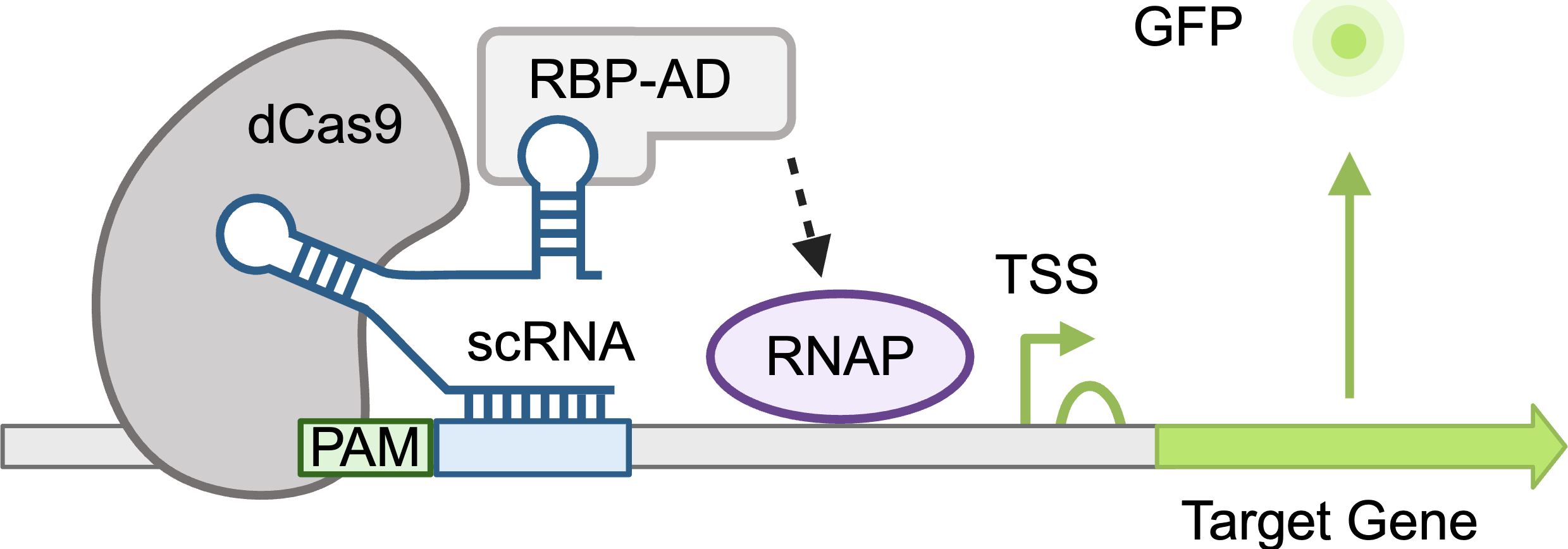}
\caption{%
  Example with $m=3$. CRISPRa (CRISPR activator) complex consists of a
  scaffold RNA (scRNA) which helps recruit the repressor to a region
  upstream of transcription start site of a target gene to be
  activated (in the illustration, a GFP-expressing gene).  The ligands are:
  an RNA-binding protein fused to an activation domain (RBP-AD),
  catalytically inactive Cas9 (dCas9),
  and the target binding site on DNA for the specific scRNA being used.
  Figure reproduced from~\cite{26_activator_repressor_journal}.}
\label{fig:scRNA}
\end{figure}

Our second motivation arose from the study of dCas9-based CRISPR
activator genetic
circuits~\cite{ManojDelVecchio2022CDC,24cdc_activator_repressor,26_activator_repressor_journal}.
In this synthetic biology construction, $m=3$.
The scaffold is a scaffold RNA (scRNA), which binds
to an RNA-binding protein, a catalytically inactive Cas9, and a
promoter region for a target gene that one wishes to activate.
See Figure~\ref{fig:scRNA} for an illustration.

\begin{figure}[ht]
\minp{0.33}{\pic{0.12}{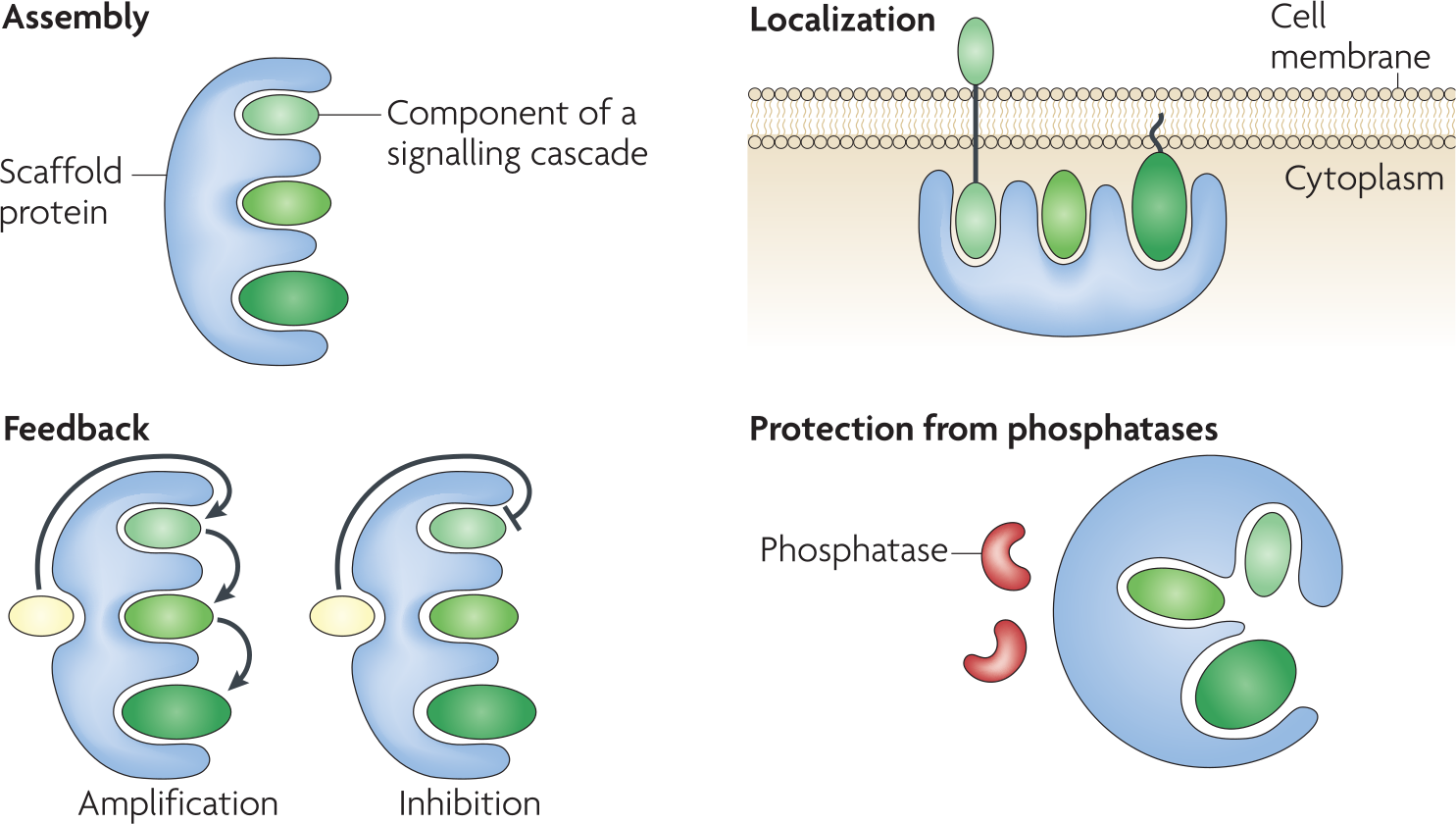}}%
\minp{0.33}{\picc{0.12}{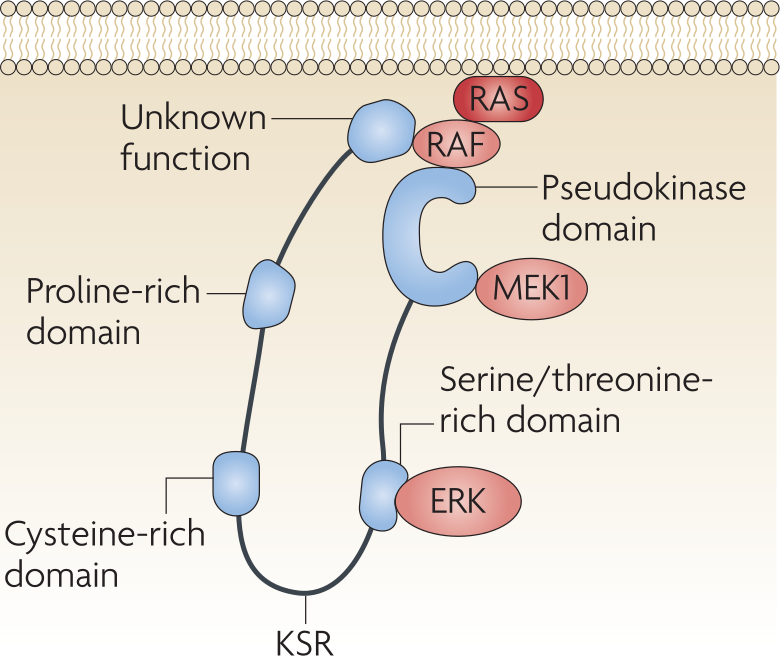}}%
\minp{0.33}{\picc{0.1}{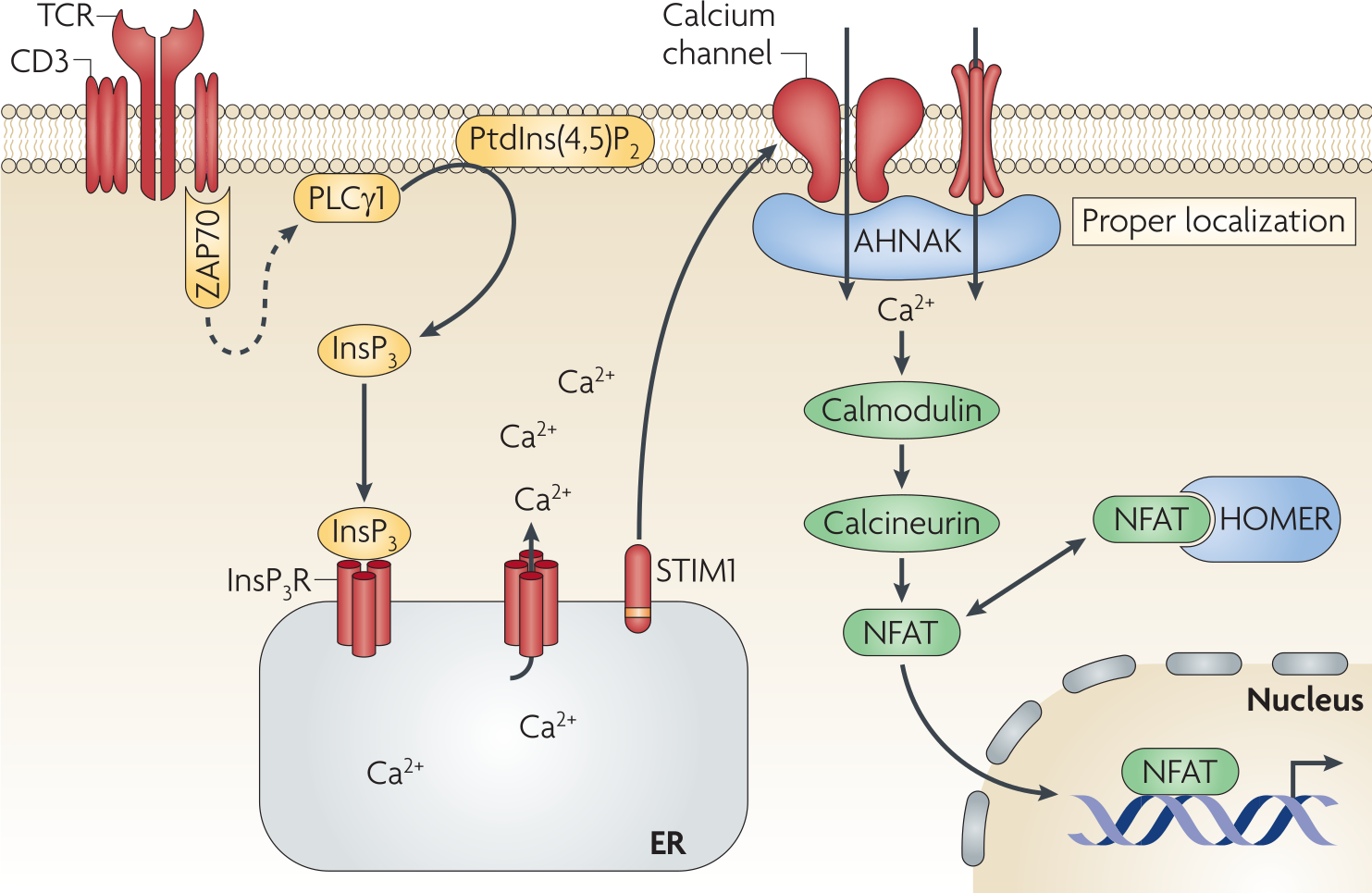}}
\caption{%
Left: In signaling pathways, scaffold proteins help assemble components and
localize them to a specific intracellular location, help facilitate
feedback loops, and protect proteins from deactivation by phosphatases. 
Middle:
In mitogen-activated protein kinase (MAPK) signalling cascades,
the scaffold protein kinase suppressor of RAS (KSR) helps
assemble a complex to enhance ERK activation.
Right:
Scaffold protein AHNAK1 plays an important role in immune-cell signalling,
leading to the release of Ca2+ from the ER after T-cell receptor (TCR)
stimulation,
by helping to localize Ca2+ channels to the plasma membrane.
Figures and discussion adapted from~\cite{ShawFilbert2009}.}
\label{fig:shaw}
\end{figure}

\begin{figure}[ht]
\minp{0.33}{\pic{0.25}{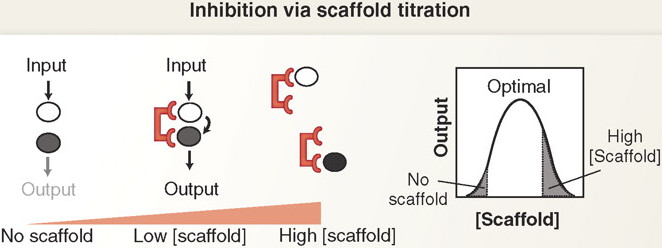}}
\minp{0.33}{\picc{0.22}{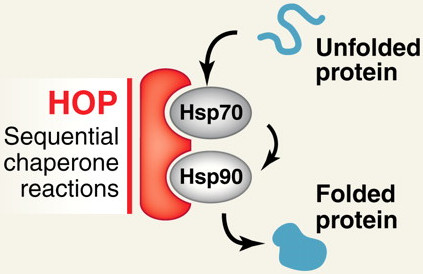}}%
\minp{0.33}{\picc{0.2}{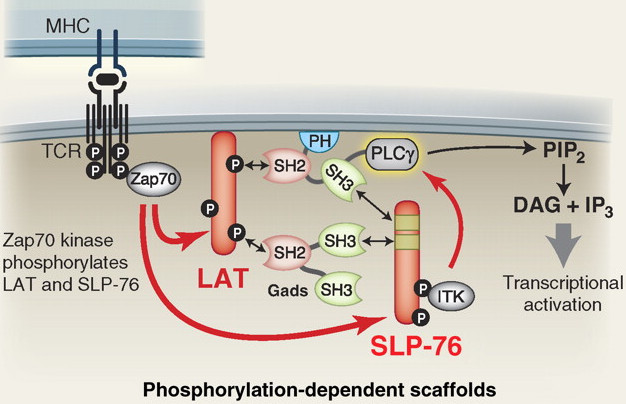}}
\caption{%
Scaffolding as controllers of cellular information transmission.
Left: As discussed earlier, at high concentrations, scaffolds may
titrate enzyme and substrate away from one another, diminishing the
effectiveness of the reactions
Middle: Scaffolding proteins play a role in assembly-line processes such as
protein folding; for example, HOP helps guide unfolded proteins
between Hsp70 and Hsp90 chaperones. 
Right:
Scaffolds can be controlled: activation of T cells leads to the
activation of two scaffolds, LAT and Slp76.
Figures adapted from~\cite{GoodZalatanLim2011}.}
\label{fig:lim}
\end{figure}

Scaffolds play a role in a wide variety of biological processes in
addition to the above two examples of conjugate antibodies and CRISPRa
synthetic biology constructs.
Especially in enzymatic systems, scaffolds can increase reaction rates by tethering sequential enzymes in close proximity, shield reactions from competing cellular processes, prevent loss or degradation of intermediates, and impose tight stoichiometry by controlling the ratio of enzymes within a cascade.
Figures~\ref{fig:shaw} and~\ref{fig:lim} provide examples.

\section{Definition of the network and stoichiometry}
\label{sec:preliminaries}

Fix an integer $m\geq 1$, and write
\[
[m]:=\{1,\ldots,m\}.
\]
We consider a scaffold species $Y$ and $m$ ligands
\[
A_1,\ldots,A_m.
\]
For each subset $I\subseteq [m]$, we denote by $Y_I$ the species consisting
of the scaffold $Y$ bound to precisely the ligands indexed by $I$. Thus
\[
Y_\varnothing=Y,
\]
and, for example, when $m=3$ one has
\[
Y_{\{1,2\}}=YA_1A_2,
\qquad
Y_{\{1,2,3\}}=YA_1A_2A_3.
\]
The species set is therefore
\be\label{eq:list_species_general}
{\mathcal S}
=\{A_i: i\in[m]\}\cup \{Y_I:I\subseteq[m]\}.
\ee
Thus the network has $m+2^m$ species. We call $Y_\varnothing$ the free scaffold, and
we call the species $Y_I$ with $I\neq\varnothing$ the scaffold-ligand
complexes. We use lower case letters for concentrations: $a_i$ denotes
the concentration of $A_i$, and $y_I$ denotes the concentration of
$Y_I$. In particular, $y_\varnothing$ is the concentration of the free
scaffold.

The reactions are the reversible binding and unbinding reactions
\be\label{eq:general_reactions}
Y_I+A_i
\;\xrightleftharpoons[k_{\mathrm{off},i}]{k_{\mathrm{on},i}}\;
Y_{I\cup\{i\}},
\qquad
I\subseteq[m],\quad i\notin I.
\ee
The key assumption is independent binding: the constants
$k_{\mathrm{on},i}$ and $k_{\mathrm{off},i}$ depend only on the ligand
$i$, and not on the subset $I$ of ligands already bound to the scaffold.
There are
\[
\sum_{I\subseteq[m]}(m-|I|)
\]
reversible pairs.
To  evaluate this sum, we group the subsets by their size $k$, where $k = |I|$. 
The number of subsets of size $k$ is given by the binomial coefficient
$\binom{m}{k}$.
Rewriting the sum in terms of $k$, we get:
\[
\sum_{k=0}^{m} \binom{m}{k} (m - k) \;=\;m \sum_{k=0}^{m} \binom{m}{k} - \sum_{k=0}^{m} k \binom{m}{k}.
\]
Using the standard binomial identity $\sum_{k=0}^{m} \binom{m}{k} = 2^m$, the first half simplifies to $m 2^m$.
We can also simplify $\sum k \binom{m}{k}$ using the identity $k \binom{m}{k} = m \binom{m-1}{k-1}$ for $k \ge 1$:
    \[
    \sum_{k=1}^{m} k \binom{m}{k} = \sum_{k=1}^{m} m \binom{m-1}{k-1} = m \sum_{j=0}^{m-1} \binom{m-1}{j} = m 2^{m-1}.
    \]
Subtracting the evaluated second term from the first gives us that the
number of reversible reaction pairs is:
\[
m 2^m - m 2^{m-1} = m(2^m - 2^{m-1}) = m 2^{m-1}.
\]
If each direction is counted separately, we have $2\times m 2^{m-1}=m2^m$ directed reactions.
Let $e_S$ denote the unit vector corresponding to the species $S$. For
the forward reaction in \eqref{eq:general_reactions}, define the
stoichiometric vector
\[
\nu_{I,i}:=e_{Y_{I\cup\{i\}}}-e_{Y_I}-e_{A_i}.
\]
Then the mass-action system can be written compactly as
\be\label{eq:general_ode}
\dot x
=
\sum_{I\subseteq[m]}\sum_{i\notin I}
\nu_{I,i}
\left(
 k_{\mathrm{on},i} y_Ia_i
 -k_{\mathrm{off},i}y_{I\cup\{i\}}
\right).
\ee

The conservation laws are the scaffold conservation law
\be\label{eq:Ytot_general}
Y_{\mathrm{tot}}
=
\sum_{I\subseteq[m]} y_I,
\ee
and the $m$ ligand conservation laws
\be\label{eq:Aitot_general}
A_{i,\mathrm{tot}}
=
a_i+
\sum_{I\subseteq[m]:\, i\in I} y_I,
\qquad i=1,\ldots,m.
\ee
Indeed, every reaction preserves the number of scaffold moieties and the
number of each ligand moiety.

For completeness, let us note the rank count. The stoichiometric
subspace is contained in the common kernel of the $m+1$ linear
conservation laws \eqref{eq:Ytot_general}--\eqref{eq:Aitot_general};
therefore its dimension is at most
\[
(m+2^m)-(m+1)=2^m-1.
\]
Conversely, choose for each nonempty subset $J\subseteq[m]$ one element
$i(J)\in J$, and consider the reaction
\[
Y_{J\setminus\{i(J)\}}+A_{i(J)}\to Y_J.
\]
The $2^m-1$ stoichiometric vectors of these reactions are linearly
independent: after projection onto the coordinates $\{Y_I:I\subseteq[m]\}$,
they are the incidence vectors of the edges of a spanning tree of the
$m$-dimensional hypercube, rooted at $\varnothing$. Hence the
stoichiometric subspace has dimension $2^m-1$, and the conservation
laws above span the left nullspace of the stoichiometry matrix.

\section{Global convergence}
\label{sec:dynamics}

Define
\[
K_i:=\frac{k_{\mathrm{on},i}}{k_{\mathrm{off},i}},
\qquad i=1,\ldots,m.
\]
We now show that the network is detailed balanced. Let
$y_\varnothing>0$ and $a_1,
\ldots,a_m>0$ be arbitrary, and define
\be\label{eq:complex_formula_general}
y_I
=
y_\varnothing\prod_{j\in I}K_j a_j,
\qquad I\subseteq[m].
\ee
Then, for every reaction \eqref{eq:general_reactions},
\[
k_{\mathrm{on},i}y_Ia_i
=
k_{\mathrm{on},i}y_\varnothing
\left(\prod_{j\in I}K_ja_j\right)a_i
=
k_{\mathrm{off},i}y_\varnothing
\left(\prod_{j\in I\cup\{i\}}K_ja_j\right)
=
k_{\mathrm{off},i}y_{I\cup\{i\}}.
\]
Thus every reversible pair is individually balanced. This proves that
our mass-action system is detailed balanced.

It follows from the standard theory of detailed-balanced mass-action
systems, see for example Theorems 14.2.1 and 14.2.3 in Feinberg's
textbook~\cite{Feinberg2019FCRNT}, that each positive stoichiometric
compatibility class contains precisely one positive equilibrium, and
that this equilibrium is locally asymptotically stable relative to its
stoichiometric compatibility class.

At any positive detailed-balanced steady state, equation
\eqref{eq:complex_formula_general} holds. Substituting this expression
into the conservation laws gives the steady-state equations
\be\label{eq:Ytot_product_general}
Y_{\mathrm{tot}}
=
y_\varnothing\prod_{j=1}^m(1+K_ja_j),
\ee
and, for each $i=1,\ldots,m$,
\be\label{eq:Aitot_product_general}
A_{i,\mathrm{tot}}
=
a_i\left(
1+K_i y_\varnothing\prod_{j\neq i}(1+K_ja_j)
\right).
\ee
Indeed,
\[
\sum_{I\ni i}y_I
=
K_i a_i y_\varnothing
\sum_{J\subseteq[m]\setminus\{i\}}\prod_{j\in J}K_ja_j
=
K_i a_i y_\varnothing\prod_{j\neq i}(1+K_ja_j).
\]

Next recall that a subset of species $\Sigma\subseteq\mathcal S$ is a
\emph{siphon} if every reaction that produces a species in $\Sigma$
consumes at least one species in $\Sigma$:
\[
\forall\, y\to y'\ \text{with}\ \supp(y')\cap\Sigma\neq\varnothing:
\quad
\supp(y)\cap\Sigma\neq\varnothing.
\]
A siphon is \emph{critical}~\cite{persistencePetri} if it does not
contain the support of any nonzero nonnegative conservation law
(P-semiflow).

Define the moiety supports
\[
S_Y:=\{Y_I:I\subseteq[m]\}
\]
and, for $i=1,\ldots,m$,
\[
S_i:=\{A_i\}\cup\{Y_I:I\subseteq[m],\ i\in I\}.
\]
The indicator vectors of these sets are P-semiflows, because every
reaction preserves the number of scaffold moieties and the number of
each ligand moiety. Their corresponding conservation laws are precisely
\eqref{eq:Ytot_general} and \eqref{eq:Aitot_general}.

\begin{lemma}\label{lem:general_siphon_contains_moiety}
Every nonempty siphon contains one of the moiety supports
\[
S_Y,S_1,\ldots,S_m.
\]
Consequently, the network has no nonempty critical siphons.
\end{lemma}

\begin{proof}
Let $\Sigma$ be a nonempty siphon.

First suppose that $A_i\in\Sigma$ for some $i$. For every
$J\subseteq[m]$ with $i\in J$, the reverse reaction
\[
Y_J\to Y_{J\setminus\{i\}}+A_i
\]
produces $A_i$. Since $A_i\in\Sigma$ and $\Sigma$ is a siphon, the
reactant $Y_J$ must belong to $\Sigma$. Therefore
\[
\{Y_J:i\in J\}\subseteq\Sigma.
\]
Together with $A_i\in\Sigma$, this gives
\[
S_i\subseteq\Sigma.
\]

It remains to consider the case in which $\Sigma$ contains none of the free
ligands. Since $\Sigma$ is nonempty, it contains at least one scaffold
species $Y_I$. Choose such an $I$ with minimal cardinality. If
$I\neq\varnothing$, choose $i\in I$. The forward reaction
\[
Y_{I\setminus\{i\}}+A_i\to Y_I
\]
produces $Y_I$. Since $A_i\notin\Sigma$ by assumption and $\Sigma$ is a
siphon, we must have
\[
Y_{I\setminus\{i\}}\in\Sigma,
\]
contradicting the minimality of $I$. Hence $I=\varnothing$, so
$Y_\varnothing\in\Sigma$.

We now show that every $Y_J$ belongs to $\Sigma$. Suppose $Y_I\in\Sigma$
and $i\notin I$. The reverse reaction
\[
Y_{I\cup\{i\}}\to Y_I+A_i
\]
produces $Y_I$. Since $\Sigma$ is a siphon, its reactant
$Y_{I\cup\{i\}}$ must belong to $\Sigma$. Starting from
$Y_\varnothing\in\Sigma$ and adding ligands one at a time, we obtain
$Y_J\in\Sigma$ for every $J\subseteq[m]$. Thus
\[
S_Y\subseteq\Sigma.
\]

In all cases, $\Sigma$ contains the support of a P-semiflow. Therefore
there are no nonempty critical siphons.
\end{proof}

\begin{theorem}[Global stability]
\label{theo:global_general}
The unique positive steady state in each positive stoichiometric
compatibility class is globally asymptotically stable with respect to
that class.
\end{theorem}

\begin{proof}
By Lemma~\ref{lem:general_siphon_contains_moiety}, the network has no
critical siphons. Therefore all trajectories with strictly positive
initial condition are persistent; that is, their omega-limit sets do
not meet the boundary of the positive orthant~\cite{persistencePetri}.

The system is detailed balanced, hence complex balanced, and therefore
admits the standard Horn--Jackson Lyapunov function. For complex-balanced
mass-action systems, every positive trajectory either converges to the
unique positive equilibrium in its stoichiometric compatibility class or
has an omega-limit point on the boundary; see, for example,
\cite{Tcell01,Anderson2008GAS}. Persistence rules out the boundary
alternative. Thus every positive trajectory converges to the unique
positive equilibrium in its stoichiometric compatibility class.
\end{proof}

\begin{remark}
An alternative proof would be based on showing directly that there are
no steady states in the boundary of any positive stoichiometric class,
and then applying the global stability criterion for complex-balanced
systems with no boundary equilibria~\cite{SiegelMacLean2000,Tcell01}.
\end{remark}

\section{Steady state of the fully bound complex as a function of $Y_{\mathrm{tot}}$}
\label{sec:biphasic}

We now fix the ligand totals
\[
A_{1,\mathrm{tot}},\ldots,A_{m,\mathrm{tot}}>0
\]
and study the positive steady state as a function of the scaffold total
$Y_{\mathrm{tot}}$. The fully bound complex is $Y_{[m]}$, and its
concentration is denoted by $y_{[m]}$. At a positive steady state,
\eqref{eq:complex_formula_general} gives
\[
y_{[m]}
=
y_\varnothing\prod_{j=1}^m K_j a_j.
\]
We prove that, when $m\geq 2$, this steady-state concentration is a
biphasic function of $Y_{\mathrm{tot}}$: it increases strictly up to a
single maximum and then decreases strictly.

\begin{theorem}[Biphasic response of the fully bound complex]
\label{theo:biphasic_general}
Assume $m\geq 2$. Fix positive ligand totals
\[
A_{1,\mathrm{tot}},\ldots,A_{m,\mathrm{tot}}>0.
\]
For each $X=Y_{\mathrm{tot}}>0$, let
\[
(y_\varnothing^*(X),a_1^*(X),\ldots,a_m^*(X))
\]
be the unique positive solution of the steady-state equations
\eqref{eq:Ytot_product_general}--\eqref{eq:Aitot_product_general}. Define
\[
F(X):=y_{[m]}^*(X)
=
y_\varnothing^*(X)\prod_{j=1}^m K_j a_j^*(X).
\]
Then $F$ is differentiable on $(0,\infty)$ and has a unique maximizer
$X_*\in(0,\infty)$. More precisely,
\[
F'(X)>0 \quad \text{for }0<X<X_*,
\qquad
F'(X)<0 \quad \text{for }X>X_*.
\]
\end{theorem}

\begin{proof}
For notational simplicity, we will omit the stars and write simply $y_{[m]}$, etc.
  Set
\[
U_i:=1+K_i a_i,
\qquad i=1,\ldots,m,
\]
so that $U_i>1$. Also write
\[
X:=Y_{\mathrm{tot}}
\]
and introduce the positive constants
\[
\alpha_i:=K_i A_{i,\mathrm{tot}},
\qquad i=1,\ldots,m.
\]
The first steady-state conservation equation gives
\[
X
=
y_\varnothing\prod_{j=1}^m U_j,
\]
and therefore
\[
y_\varnothing=
\frac{X}{\prod_{j=1}^m U_j}.
\]
For the $i$th ligand conservation equation, using
\[
a_i=\frac{U_i-1}{K_i}
\]
and
\[
y_\varnothing\prod_{j\neq i}U_j=\frac{X}{U_i},
\]
we obtain
\[
A_{i,\mathrm{tot}}
=
\frac{U_i-1}{K_i}
\left(1+\frac{K_iX}{U_i}\right).
\]
Multiplying by $K_i$ gives
\be\label{eq:Ui_equation_general}
\frac{(U_i-1)(U_i+K_iX)}{U_i}=\alpha_i.
\ee
Equivalently,
\be\label{eq:Ui_quadratic_general}
U_i^2+(K_iX-\alpha_i-1)U_i-K_iX=0.
\ee

We first list some elementary consequences of this equation. For fixed
$X>0$, let
\[
g_i(U):=U^2+(K_iX-\alpha_i-1)U-K_iX.
\]
Then
\[
g_i(0)=-K_iX<0,
\qquad
\lim_{U\to\infty}g_i(U)=+\infty.
\]
Thus $g_i$ has a positive root. Since the product of the two roots of
\eqref{eq:Ui_quadratic_general} is $-K_iX<0$, this positive root is
unique. Moreover,
\[
g_i(1)=1+K_iX-\alpha_i-1-K_iX=-\alpha_i<0.
\]
Because the parabola opens upward, the unique positive root satisfies
$U_i>1$.

We also note that this positive root depends smoothly on $X$. Consider
\[
g_i(U,X):=U^2+(K_iX-\alpha_i-1)U-K_iX
\]
as a function of $(U,X)$. At a root,
\[
K_iX-\alpha_i-1=\frac{K_iX}{U}-U,
\]
and hence
\[
\frac{\partial g_i}{\partial U}
=
2U+K_iX-\alpha_i-1
=
U+\frac{K_iX}{U}>0.
\]
The implicit function theorem therefore gives a smooth function
\[
U_i=U_i(X)>1,
\qquad X>0.
\]
This applies independently for each $i=1,\ldots,m$.

Conversely, once the functions $U_i(X)$ have been determined, define
\[
a_i(X):=\frac{U_i(X)-1}{K_i},
\qquad
 y_\varnothing(X):=\frac{X}{\prod_{j=1}^m U_j(X)}.
\]
Reversing the preceding algebra shows that these quantities satisfy the
steady-state conservation equations
\eqref{eq:Ytot_product_general}--\eqref{eq:Aitot_product_general}. Thus
we have recovered the unique positive steady state for the prescribed
totals.

Now the concentration of the fully bound complex is
\[
F(X)=y_{[m]}^*(X)
=
y_\varnothing(X)\prod_{j=1}^m K_j a_j(X).
\]
Using $K_i a_i=U_i-1$ and
$y_\varnothing=X/\prod_j U_j$, this becomes
\[
F(X)
=
X\prod_{j=1}^m\frac{U_j-1}{U_j}.
\]
From \eqref{eq:Ui_equation_general},
\[
\frac{U_i-1}{U_i}
=
\frac{\alpha_i}{U_i+K_iX}.
\]
Therefore
\be\label{eq:F_formula_general}
F(X)
=
\frac{\left(\prod_{j=1}^m\alpha_j\right)X}
{\prod_{j=1}^m(U_j+K_jX)}.
\ee
Since the functions $U_i(X)$ are smooth, $F$ is differentiable.

We now compute the derivative. Since smoothness has already been
established, we may differentiate \eqref{eq:Ui_equation_general}, or
equivalently its form
\[
U_i+K_iX-1-\frac{K_iX}{U_i}=\alpha_i.
\]
This gives
\[
U_i'
+
K_i
-
\frac{K_i}{U_i}
+
\frac{K_iX}{U_i^2}U_i'
=0,
\]
and hence
\be\label{eq:Ui_prime_general}
U_i'
=
-\frac{K_iU_i(U_i-1)}{U_i^2+K_iX}<0.
\ee
Furthermore,
\[
U_i'+K_i
=
K_i\frac{U_i+K_iX}{U_i^2+K_iX}.
\]
Thus
\be\label{eq:log_factor_derivative_general}
\frac{d}{dX}\log(U_i+K_iX)
=
\frac{K_i}{U_i^2+K_iX}.
\ee
Taking logarithmic derivatives in \eqref{eq:F_formula_general} and using
\eqref{eq:log_factor_derivative_general}, we obtain
\[
\frac{d}{dX}\log F
=
\frac1X-
\sum_{i=1}^m\frac{K_i}{U_i^2+K_iX}.
\]
Equivalently,
\be\label{eq:logF_derivative_general}
\frac{d}{dX}\log F
=
\frac{1-S(X)}{X},
\ee
where
\[
S(X):=\sum_{i=1}^m T_i(X),
\qquad
T_i(X):=\frac{K_iX}{U_i(X)^2+K_iX}.
\]

We claim that each $T_i$ is strictly increasing. Indeed, differentiating
and using \eqref{eq:Ui_prime_general},
\[
T_i'(X)
=
\frac{
K_i(U_i^2+K_iX)-K_iX(2U_iU_i'+K_i)
}{(U_i^2+K_iX)^2}.
\]
The $K_i^2X$ terms cancel, so
\[
T_i'(X)
=
\frac{K_iU_i^2-2K_iXU_iU_i'}{(U_i^2+K_iX)^2}>0,
\]
because $K_i>0$, $X>0$, $U_i>0$, and $U_i'<0$. Hence $S$ is strictly
increasing on $(0,\infty)$.

It remains to compute the endpoint limits. From
\eqref{eq:Ui_equation_general},
\[
U_i-1<\alpha_i,
\]
because $(U_i+K_iX)/U_i>1$. Thus $U_i$ is bounded. Letting
$X\to0^+$ in \eqref{eq:Ui_equation_general} gives
\[
U_i\to1+\alpha_i,
\]
and consequently
\[
T_i(X)\to0
\qquad\text{as }X\to0^+.
\]
Thus
\[
S(X)\to0
\qquad\text{as }X\to0^+.
\]

On the other hand, \eqref{eq:Ui_equation_general} gives
\[
U_i-1=
\frac{\alpha_iU_i}{U_i+K_iX}.
\]
Since $1<U_i<1+\alpha_i$, we have
\[
0<U_i-1
\leq
\frac{\alpha_i(1+\alpha_i)}{K_iX}
\to0
\qquad\text{as }X\to\infty.
\]
Therefore
\[
U_i\to1
\qquad\text{as }X\to\infty,
\]
and hence
\[
T_i(X)\to1
\qquad\text{as }X\to\infty.
\]
Thus
\[
S(X)\to m
\qquad\text{as }X\to\infty.
\]

Since $m\geq2$, the strictly increasing continuous function $S$ goes
from $0$ to $m>1$. Therefore there exists a unique $X_*\in(0,\infty)$
such that
\[
S(X_*)=1.
\]
Moreover, $S(X)<1$ for $0<X<X_*$ and $S(X)>1$ for $X>X_*$. By
\eqref{eq:logF_derivative_general}, and since $F(X)>0$, this implies
\[
F'(X)>0 \quad \text{for }0<X<X_*,
\qquad
F'(X)<0 \quad \text{for }X>X_*.
\]
Thus the steady-state concentration of the fully bound complex is
strictly increasing before $X_*$ and strictly decreasing after $X_*$. In
particular, it has a unique maximum.
\end{proof}

\begin{remark}
The restriction $m\geq2$ is essential for a finite biphasic maximum. If
$m=1$, then $S(X)=T_1(X)$ is strictly increasing from $0$ to $1$, but it
does not reach $1$ at any finite value of $X$. In that case the fully
bound complex is monotone increasing and approaches a limiting value as
$Y_{\mathrm{tot}}\to\infty$.
\end{remark}

\section{Monotonicity of the free species and partial complexes}
\label{sec:monotonefree}

In the preceding section we showed that, for fixed positive ligand totals
$A_{1,\mathrm{tot}},\ldots,A_{m,\mathrm{tot}}$
and for each positive value of the total scaffold concentration
$X:=Y_{\mathrm{tot}}$, there is a unique positive steady state, and
we proved that the fully bound complex is biphasic when $m\geq 2$.
In this section we present complementary monotonicity properties of the
free scaffold, the free ligands, and the singly bound complexes. We also
explain why these monotonicity properties, by themselves, do not imply
the biphasic behavior of the fully bound complex.

Throughout this section, \(y_\varnothing,a_1,\ldots,a_m\) denote the
unique positive steady-state concentrations associated with the fixed
ligand totals \(A_{1,\mathrm{tot}},\ldots,A_{m,\mathrm{tot}}\) and the
variable scaffold total \(X=Y_{\mathrm{tot}}\). We continue to use the
notation
\[
U_i:=1+K_i a_i,\qquad
\alpha_i:=K_iA_{i,\mathrm{tot}},
\qquad i=1,\ldots,m.
\]
At steady state,
\[
X=y_\varnothing\prod_{j=1}^m U_j
\]
and
\[
\frac{(U_i-1)(U_i+K_iX)}{U_i}=\alpha_i,
\qquad i=1,\ldots,m.
\]
Equivalently,
\[
U_i+K_iX-1-\frac{K_iX}{U_i}=\alpha_i.
\]

\begin{theorem}
\label{theo:monotonefree}
Fix positive ligand totals
\[
A_{1,\mathrm{tot}},\ldots,A_{m,\mathrm{tot}}.
\]
At the unique positive steady state, the free scaffold concentration
\(y_\varnothing^{\maybestar}\) is a strictly increasing function of
\(X=Y_{\mathrm{tot}}\), and each free ligand concentration
\(a_i^{\maybestar}\) is a strictly decreasing function of \(X\). In
contrast to the latter, each singly bound complex
\[
y_{\{i\}}^{\maybestar}=K_i y_\varnothing^{\maybestar}a_i^{\maybestar}
\]
is a strictly increasing function of \(X\). Moreover,
\[
\lim_{X\to0^+}y_\varnothing^{\maybestar}(X)=0,
\qquad
\lim_{X\to\infty}y_\varnothing^{\maybestar}(X)=\infty,
\]
\[
\lim_{X\to0^+}a_i^{\maybestar}(X)=A_{i,\mathrm{tot}},
\qquad
\lim_{X\to\infty}a_i^{\maybestar}(X)=0,
\]
and
\[
\lim_{X\to0^+}y_{\{i\}}^{\maybestar}(X)=0,
\qquad
\lim_{X\to\infty}y_{\{i\}}^{\maybestar}(X)=A_{i,\mathrm{tot}}.
\]
\end{theorem}

\begin{proof}
The functions \(U_i=U_i(X)\) are smooth by the implicit-function
argument used above. Differentiating
\[
U_i+K_iX-1-\frac{K_iX}{U_i}=\alpha_i
\]
with respect to \(X\) gives
\[
U_i'
+
K_i
-
\frac{K_i}{U_i}
+
\frac{K_iX}{U_i^2}U_i'
=
0.
\]
Hence
\[
U_i'
=
-\frac{K_iU_i(U_i-1)}{U_i^2+K_iX}<0.
\]
Since
\[
a_i=\frac{U_i-1}{K_i},
\]
we obtain
\[
\frac{da_i^{\maybestar}}{dX}
=
\frac{U_i'}{K_i}
=
-\frac{U_i(U_i-1)}{U_i^2+K_iX}<0.
\]
Thus each free ligand concentration is strictly decreasing as a function
of \(X\).

On the other hand,
\[
y_\varnothing^{\maybestar}(X)=\frac{X}{\prod_{j=1}^m U_j(X)}.
\]
Taking logarithmic derivatives gives
\[
\frac{d}{dX}\log y_\varnothing^{\maybestar}
=
\frac1X-\sum_{j=1}^m\frac{U_j'}{U_j}.
\]
Using the formula for \(U_j'\), this becomes
\[
\frac{d}{dX}\log y_\varnothing^{\maybestar}
=
\frac1X+
\sum_{j=1}^m
\frac{K_j(U_j-1)}{U_j^2+K_jX}>0.
\]
Since \(y_\varnothing^{\maybestar}>0\), it follows that
\[
\frac{dy_\varnothing^{\maybestar}}{dX}>0.
\]
Thus the free scaffold concentration is strictly increasing as a
function of \(X=Y_{\mathrm{tot}}\).

Next consider a singly bound complex. Since
\[
y_{\{i\}}^{\maybestar}
=
K_i y_\varnothing^{\maybestar}a_i^{\maybestar}
=
X\frac{U_i-1}{\prod_{j=1}^m U_j},
\]
we have
\[
\frac{d}{dX}\log y_{\{i\}}^{\maybestar}
=
\frac1X
+
\frac{U_i'}{U_i-1}
-
\sum_{j=1}^m\frac{U_j'}{U_j}.
\]
Equivalently,
\[
\frac{d}{dX}\log y_{\{i\}}^{\maybestar}
=
\frac1X
+
\left(
\frac{U_i'}{U_i-1}-\frac{U_i'}{U_i}
\right)
-
\sum_{j\ne i}\frac{U_j'}{U_j}.
\]
Using the formula for \(U_j'\), we obtain
\[
\frac{d}{dX}\log y_{\{i\}}^{\maybestar}
=
\frac1X
-
\frac{K_i}{U_i^2+K_iX}
+
\sum_{j\ne i}
\frac{K_j(U_j-1)}{U_j^2+K_jX}.
\]
The first two terms satisfy
\[
\frac1X-\frac{K_i}{U_i^2+K_iX}
=
\frac{U_i^2}{X(U_i^2+K_iX)}>0,
\]
and every term in the remaining sum is positive. Hence
\[
\frac{d}{dX}\log y_{\{i\}}^{\maybestar}>0.
\]
Since \(y_{\{i\}}^{\maybestar}>0\), this proves that each singly bound
complex is strictly increasing as a function of \(X\).

It remains to check the endpoint behavior. From
\[
\frac{(U_i-1)(U_i+K_iX)}{U_i}=\alpha_i
\]
we have
\[
U_i-1<\alpha_i,
\]
because
\[
\frac{U_i+K_iX}{U_i}>1.
\]
Thus \(U_i\) remains bounded as \(X\to0^+\). Letting \(X\to0^+\) in the
defining equation gives
\[
U_i\to1+\alpha_i.
\]
Therefore
\[
a_i^{\maybestar}(X)=\frac{U_i(X)-1}{K_i}\to\frac{\alpha_i}{K_i}
=
A_{i,\mathrm{tot}},
\]
and
\[
y_\varnothing^{\maybestar}(X)=\frac{X}{\prod_{j=1}^m U_j(X)}\to0.
\]
It follows also that
\[
y_{\{i\}}^{\maybestar}(X)
=
X\frac{U_i(X)-1}{\prod_{j=1}^m U_j(X)}
\to0.
\]

For the limit as \(X\to\infty\), rewrite the defining equation as
\[
U_i-1=\frac{\alpha_iU_i}{U_i+K_iX}.
\]
Since \(1<U_i<1+\alpha_i\), we obtain
\[
0<U_i-1
\le
\frac{\alpha_i(1+\alpha_i)}{K_iX}\to0.
\]
Hence
\[
U_i\to1,
\]
and therefore
\[
a_i^{\maybestar}(X)=\frac{U_i(X)-1}{K_i}\to0.
\]
Since \(U_j(X)\to1\) for every \(j\), we also have
\[
y_\varnothing^{\maybestar}(X)
=
\frac{X}{\prod_{j=1}^m U_j(X)}
\to\infty.
\]
Finally,
\[
X(U_i-1)
=
\frac{\alpha_iXU_i}{U_i+K_iX}
\to
\frac{\alpha_i}{K_i}
=
A_{i,\mathrm{tot}},
\]
and hence
\[
y_{\{i\}}^{\maybestar}(X)
=
X\frac{U_i(X)-1}{\prod_{j=1}^m U_j(X)}
\to A_{i,\mathrm{tot}}.
\]
This proves the theorem.
\end{proof}

The preceding theorem gives a useful qualitative picture. As the total
amount of scaffold is increased, the free scaffold concentration
increases, while the free concentrations of all ligands decrease. At the
same time, the amount of each ligand captured somewhere on a scaffold,
represented by the singly bound complex \(y_{\{i\}}\), increases and approaches the total amount of that
ligand. Thus a singly bound complex behaves like a capture curve: adding
more scaffold can only increase the amount of ligand \(i\) that is bound
to some scaffold molecule.

\subsection{Higher partially bound complexes}

The behavior of complexes bound to two or more ligands is subtler. For
\(I\subseteq\{1,\ldots,m\}\), recall that
\[
y_I
=
\left(\prod_{i\in I}K_i\right)y_\varnothing\prod_{i\in I}a_i.
\]
Equivalently,
\[
y_I
=
X\frac{\prod_{i\in I}(U_i-1)}{\prod_{j=1}^m U_j}.
\]
Taking a logarithmic derivative gives
\[
\frac{d}{dX}\log y_I
=
\frac1X
+
\sum_{i\in I}
\left(
\frac{U_i'}{U_i-1}-\frac{U_i'}{U_i}
\right)
-
\sum_{j\notin I}\frac{U_j'}{U_j}.
\]
Using the formula for \(U_i'\), this becomes
\[
\frac{d}{dX}\log y_I
=
\frac1X
-
\sum_{i\in I}\frac{K_i}{U_i^2+K_iX}
+
\sum_{j\notin I}
\frac{K_j(U_j-1)}{U_j^2+K_jX}.
\]
Equivalently, if
\[
T_i(X):=\frac{K_iX}{U_i^2+K_iX},
\]
then
\[
X\frac{d}{dX}\log y_I
=
1
-
\sum_{i\in I}T_i(X)
+
\sum_{j\notin I}(U_j(X)-1)T_j(X).
\]

This formula shows that if \(|I|\geq2\), then \(y_I\) increases for
small \(X\) and decreases for large \(X\). Indeed, as \(X\to0^+\),
all the \(T_i(X)\) tend to zero, so
\[
X\frac{d}{dX}\log y_I\to1.
\]
On the other hand, as \(X\to\infty\), we have \(T_i(X)\to1\) and
\(U_j(X)-1\to0\). Hence
\[
X\frac{d}{dX}\log y_I\to 1-|I|.
\]
Thus, if \(|I|\ge2\), the logarithmic derivative is eventually negative.
In particular, such a partially bound complex has at least one maximum.

For the fully bound complex \(I=\{1,\ldots,m\}\), the last sum is absent,
and the argument above showed that the remaining expression changes
sign exactly once. This is the special algebraic fact responsible for the
unique biphasic response of the fully bound complex.

For a partially bound complex that is neither singly bound nor fully
bound, the formula above does not in general force uniqueness of the
maximum. For example, when \(m=3\), the doubly bound complex
\[
y_{\{1,2\}}
=
K_1K_2y_\varnothing a_1a_2
\]
satisfies
\[
y_{\{1,2\}}
=
X\frac{(U_1-1)(U_2-1)}{U_1U_2U_3},
\]
and
\[
X\frac{d}{dX}\log y_{\{1,2\}}
=
1-T_1(X)-T_2(X)+(U_3(X)-1)T_3(X).
\]
The first three terms are analogous to the terms that appear for the
fully bound complex, but the additional positive term
\[
(U_3(X)-1)T_3(X)
\]
can affect the sign of the logarithmic derivative in an intermediate
range. Thus the simple uniqueness argument available for the fully bound
complex does not automatically extend to arbitrary partially bound
complexes.

The intuition is that a singly bound complex \(y_{\{i\}}\) only asks
whether ligand \(i\) is captured by some scaffold molecule. As more
scaffold is added, this captured amount increases monotonically. A fully
bound complex asks for all ligands to be bound to the same scaffold
molecule. At low scaffold abundance, adding scaffold helps form such
complexes; at high scaffold abundance, however, different ligands tend to
be distributed among different scaffold molecules, and the fully bound
complex decreases. Intermediate partial complexes lie between these two
extremes. They must eventually decrease once they contain two or more
ligands, but they need not satisfy the same uniqueness theorem as the
fully bound complex.

\begin{remark}
The absence of a general uniqueness theorem for intermediate partial
complexes is not merely a technical limitation of the proof. 
For the doubly bound complex \(y_{\{1,2\}}\) and $m=3$, the sign of
the derivative is the sign of
\[
X\frac{d}{dX}\log y_{\{1,2\}}
=
1-T_1(X)-T_2(X)+(U_3(X)-1)T_3(X).
\]
This expression may change sign more than once; see
Figure~\ref{fig:numerical_example} for a numerical example.
\begin{figure}[ht]
\minp{0.2}{\picc{0.2}{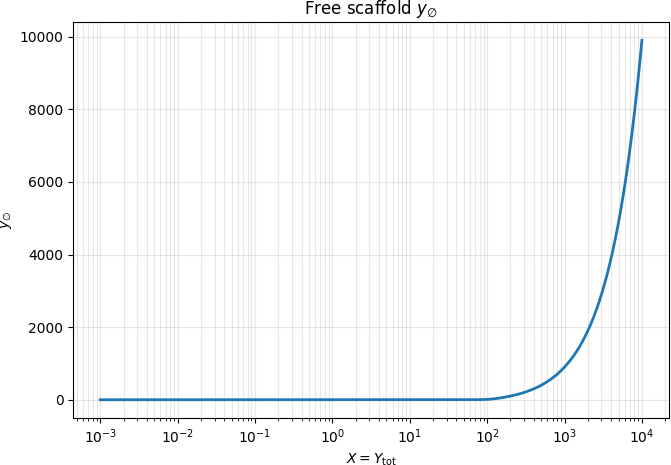}}%
\minp{0.2}{\picc{0.2}{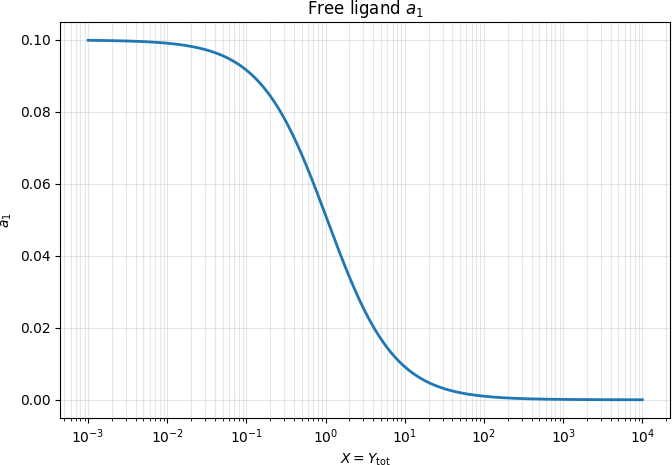}}%
\minp{0.2}{\picc{0.2}{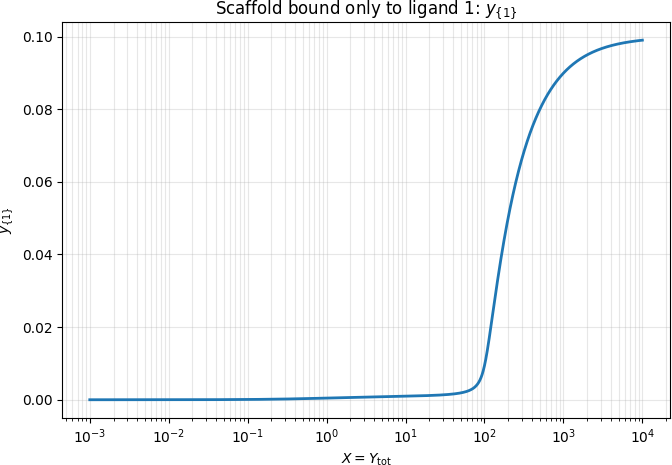}}%
\minp{0.2}{\picc{0.2}{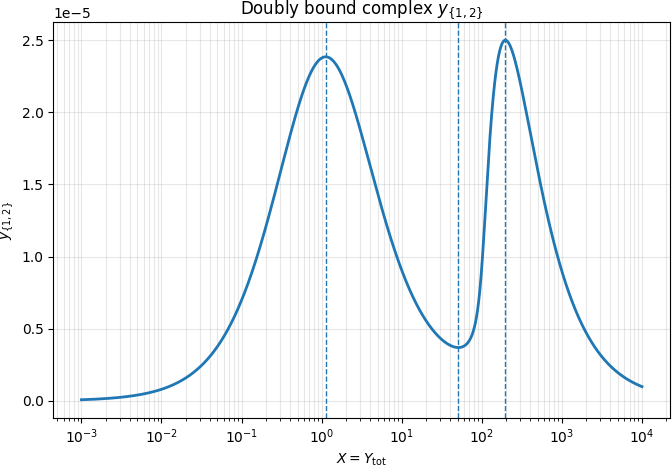}}%
\minp{0.2}{\picc{0.2}{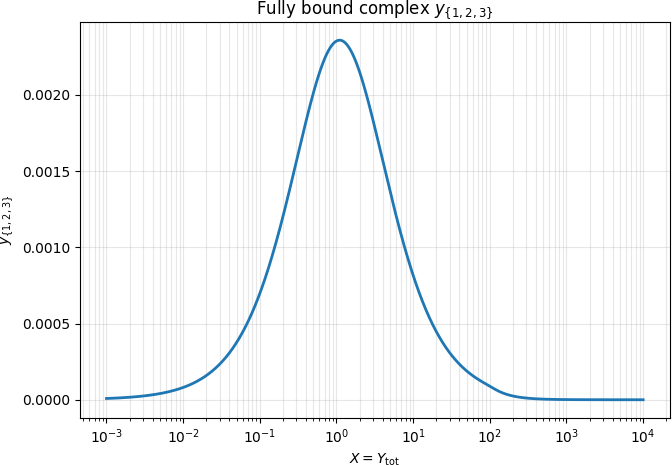}}
\caption{Numerical example. The three-ligand case with
$K_1=K_2=K_3=1$,
and
$\alpha_1=\alpha_2=0.1$, $\alpha_3=100$.
Equivalently,
$A_{1,\mathrm{tot}}=A_{2,\mathrm{tot}}=0.1$ and
$A_{3,\mathrm{tot}}=100$.
Top plots from left to right:
free scaffold $y_{\varnothing}$,
free ligand $a_1$,
scaffold bound only to ligand $a_1$,
scaffold bound to two ligands $a_1,a_2$,
and fully bound complex (biphasic).
Note that \(y_{\{1,2\}}\) has two local maxima separated by a local
minimum. The critical points are approximately at
$X\approx 1.12$, $51$, and $197$,
with the sign pattern $+,\;-,\;+,\;-$
for the scaled logarithmic derivative $X\frac{d}{dX}\log y_{\{1,2\}}$.
}
\label{fig:numerical_example}
\end{figure}
This example shows that, already for three ligands, a doubly bound complex
is not guaranteed to be monotone or uniquely biphasic.
\end{remark}

\begin{remark}
Let us discuss why, in the above example, different partially bound
complexes have qualitatively different dependence on \(Y_{\mathrm{tot}}\).
In the case \(m=3\), the logarithmic derivative of
a doubly bound complex has the form
\[
X\frac{d}{dX}\log y_{\{i,j\}}
=
1-T_i(X)-T_j(X)+(U_k(X)-1)T_k(X),
\]
where \(k\) is the index of the ligand not present in the complex, and
\[
T_\ell(X):=\frac{K_\ell X}{U_\ell(X)^2+K_\ell X}.
\]
The terms $1-T_i(X)-T_j(X)$
are the terms that drive the initial increase followed by an
eventual decrease. Indeed, \(T_i,T_j\to0\) as \(X\to0^+\), while
\(T_i,T_j\to1\) as \(X\to\infty\). The additional term
$(U_k(X)-1)T_k(X)$
comes from the ligand not included in the complex. This term is always
positive, and it can create an additional upward contribution to the
logarithmic derivative over an intermediate range of \(X\).
For the previous example, see Figure~\ref{fig:numerical_example_extra}
for plots of steady states for the remaining species.
(In this very special example, since we picked $K_1=K_2$ and
$A_{1,\mathrm{tot}}=A_{2,\mathrm{tot}}$, by symmetry the solutions
for ligands 1 and 2 are the same.)
\begin{figure}[ht]
\minp{0.33}{\picc{0.25}{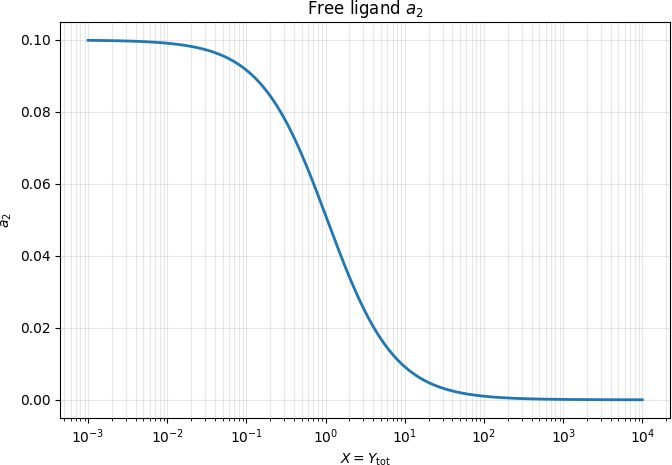}}%
\minp{0.33}{\picc{0.25}{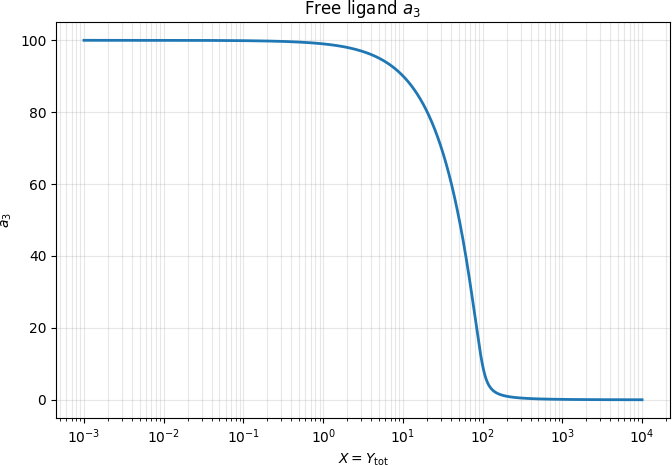}}%
\minp{0.33}{\picc{0.25}{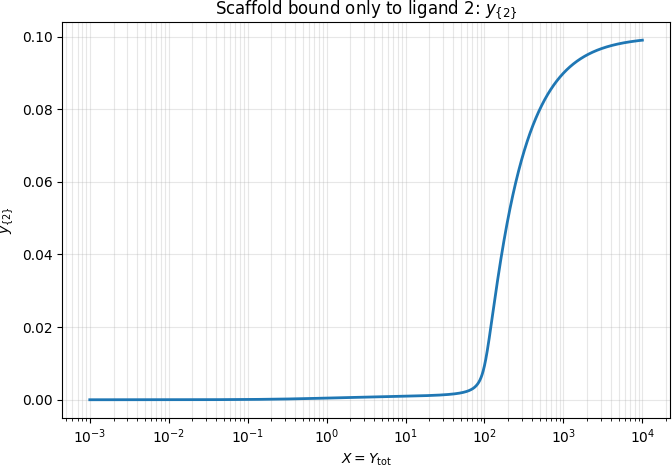}}

\minp{0.33}{\picc{0.25}{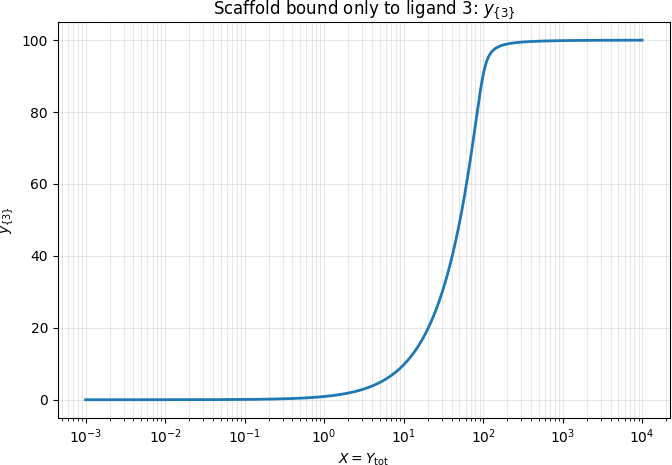}}%
\minp{0.33}{\picc{0.25}{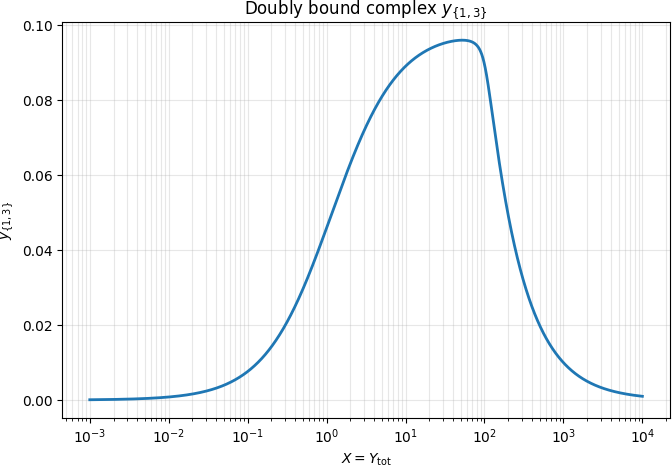}}%
\minp{0.33}{\picc{0.25}{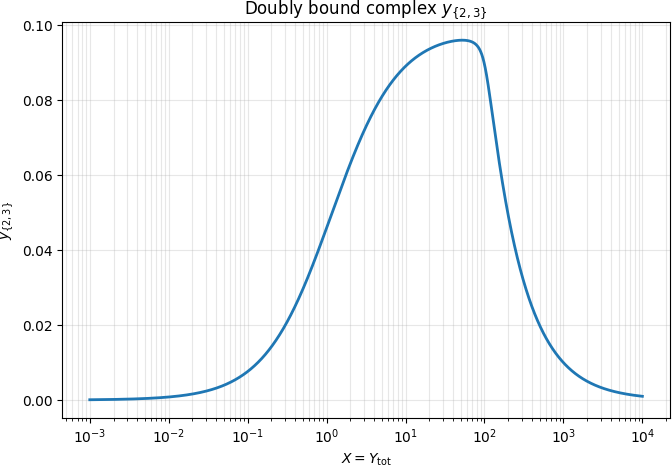}}
\caption{More plots for the numerical example considered earlier.
The plots show the free ligands $a_2$ and $a_3$ (both monotonically decreasing);
scaffold bound only to ligand $a_2$ or $a_3$ (both monotonically increasing);
scaffold bound to two ligands $a_1,a_3$ and
scaffold bound to two ligands $a_2,a_3$ (both biphasic).
}
\label{fig:numerical_example_extra}
\end{figure}

This example also explains why $y_{\{1,3\}}$ and $y_{\{2,3\}}$ are biphasic, whereas
$y_{\{1,2\}}$ is not.
For the complex \(y_{\{1,2\}}\), the missing ligand is ligand \(3\).
Since \(\alpha_3\) is large, the quantity \(U_3-1\), which is
proportional to the free concentration of ligand \(3\), remains large
over a broad range of intermediate \(X\)-values. Thus the positive term
\[
(U_3(X)-1)T_3(X)
\]
is large enough to make the logarithmic derivative of \(y_{\{1,2\}}\)
positive again after it has already become negative. Numerically, the
logarithmic derivative changes sign three times, as shown above.
Thus \(y_{\{1,2\}}\) has two local maxima separated by a local minimum.

By contrast, for \(y_{\{1,3\}}\) the missing ligand is ligand \(2\), and
for \(y_{\{2,3\}}\) the missing ligand is ligand \(1\). In the above
parameter regime these missing ligands have much smaller values of
\(\alpha_i\). The corresponding positive terms
\[
(U_2(X)-1)T_2(X)
\qquad\text{and}\qquad
(U_1(X)-1)T_1(X)
\]
are therefore not large enough to create additional sign changes. Hence
\(y_{\{1,3\}}\) and \(y_{\{2,3\}}\) are biphasic in this example, while
\(y_{\{1,2\}}\) is not. This illustrates why the
special uniqueness theorem for the fully bound complex does not extend
automatically to arbitrary partially bound complexes.
\end{remark}

\begin{remark}
For intermediate partial complexes, more than two local maxima can occur
when there are more ligands. For an example (likely of purely
mathematical interest), consider the four-ligand case
and the partial complex \(y_{\{1,2\}}\). This complex is missing two
ligands, namely ligands \(3\) and \(4\). Each missing ligand can produce
a delayed ``release'' effect analogous to the one described above for
the three-ligand case.
To see this concretely, take
\[
m=4,\qquad K_1=K_2=K_3=K_4=1,
\]
and
\[
\alpha_1=\alpha_2=0.01,\qquad
\alpha_3=10,\qquad
\alpha_4=1000.
\]
Since all \(K_i=1\), these are also the corresponding ligand totals.
For \(I=\{1,2\}\), the scaled logarithmic derivative is
\[
X\frac{d}{dX}\log y_{\{1,2\}}
=
1-T_1(X)-T_2(X)
+(U_3(X)-1)T_3(X)
+(U_4(X)-1)T_4(X).
\]
The two positive terms correspond to the two ligands missing from the
partial complex. In this example, the two missing ligands are present on
well-separated concentration scales. As the scaffold concentration is
increased, ligand \(3\) and ligand \(4\) are titrated at different
scaffold levels, and these two titration events can produce two separate
secondary increases of \(y_{\{1,2\}}\).

Numerically, the derivative changes sign at approximately
\[
X = 1.273200627, 5.936241697, 17.02106061, 495.7488044, 2002.055823.
\]
The sign pattern is
\[
+,\ -,\ +,\ -,\ +,\ -.
\]
Thus \(y_{\{1,2\}}\) has three local maxima, near
\[
X = 1.273200627, 17.02106061, X = 2002.055823.
\]
separated by two local minima.
See Figure~\ref{fig:numerical_example_three_max}.
\begin{figure}[ht]
\picc{0.4}{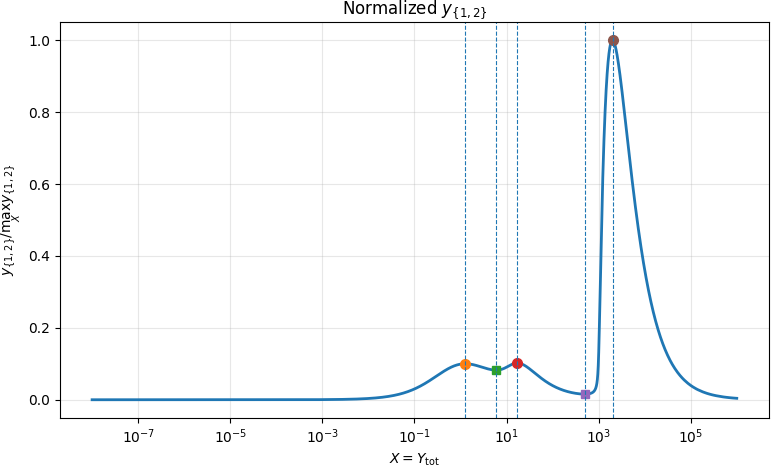}
\caption{An example with three local maxima for a partial complex
  $y_{\{1,2\}}$ when there are four ligands. The vertical axis has
  been scaled to the largest local maximum.}
\label{fig:numerical_example_three_max}
\end{figure}
This example illustrates that, while the
fully bound complex has a unique maximum for every \(m\ge2\), intermediate
partial complexes can exhibit more complicated multiphasic dose-response
behavior. Biochemically, the mechanism is that each missing abundant
ligand can first suppress the partial complex by completing it, and later
allow it to reappear once that ligand is titrated by excess scaffold.
\end{remark}

\begin{remark}
Let us now discuss more intutively (or at least attempt to rationalize!)
the origin of the repeated maxima in the intermediate complex
\(y_{\{1,2\}}\), in the above numerical example,
For a partial complex such as \(y_{\{1,2\}}\) in the four-ligand case,
three local maxima are possible. The intuition is that \(y_{\{1,2\}}\)
is missing two ligands, namely ligands \(3\) and \(4\). Each missing
ligand can create its own delayed ``release'' effect. For \(m=3\), the
complex \(y_{\{1,2\}}\) was missing only ligand \(3\), and that single
missing abundant ligand could produce one extra rise, giving two peaks.
For \(m=4\), if ligands \(3\) and \(4\) are abundant on two separated
scales, they can produce two extra rises, giving three peaks.
A useful way to think about this is by thinking of probabilities as follows:
\[
y_{\{1,2\}}
=
\text{scaffold amount}
\times
P(\text{ligands \(1\) and \(2\) both bound})
\times
P(\text{ligand \(3\) absent})
\times
P(\text{ligand \(4\) absent)}.
\]
Thus, if ligands \(1\) and \(2\) are scarce while ligands \(3\) and \(4\)
are much more abundant, the following sequence can occur as the total
scaffold concentration is increased. First, \(y_{\{1,2\}}\) rises because
more scaffold is available. It then falls because ligands \(3\) and \(4\)
tend to complete the partial complex. It rises again when ligand \(3\)
begins to be titrated by scaffold. It falls again while ligand \(4\) is
still abundant enough to complete the complex. It rises a third time when
ligand \(4\) is also titrated. Finally, it falls because ligands \(1\)
and \(2\) become diluted among many scaffold molecules.

\end{remark}

\begin{remark}
\label{rem:factors}
It is tempting to interpret a biphasic response simply as the product
of an increasing function and one or more decreasing functions. This
intuition is useful but incomplete. In general, the product of an
increasing positive function and a decreasing positive function need not
be biphasic.
In fact, products of increasing and decreasing positive
functions can exhibit essentially arbitrary positive profiles.
To see this, let \(h\) be any positive \(C^1\) function on an interval
\(I\). We will show that we can write \(h\) as a product
\[
h(x)=u(x)v(x),
\]
where \(u\) is strictly increasing and \(v\) is strictly decreasing.
First choose a continuous function \(g\) such that
\[
g(x)>\max\left\{0,\frac{h'(x)}{h(x)}\right\}
\qquad\text{for all }x\in I,
\]
and define
$G(x):=\int_{x_0}^x g(s)\,ds$
for some fixed \(x_0\in I\). Next, set
$u(x):=e^{G(x)}$ and $v(x):=h(x)e^{-G(x)}$.
Observe that $u(x)v(x)=e^{G(x)}h(x)e^{-G(x)}=h(x)$.
Since $u'(x)=g(x)e^{G(x)}>0$, we know that \(u\) is strictly increasing.
On the other hand, since
$\frac{d}{dx}\log v(x)=\frac{h'(x)}{h(x)}-g(x)<0$, we conclude that
\(v\) is strictly decreasing.
Thus, by choosing \(h\), one can obtain a product with any prescribed
positive \(C^1\) behavior: monotone increasing, monotone decreasing,
unimodal, multimodal, oscillatory, or nearly flat.
We may think of \(v\) as being almost the inverse of the
increasing factor \(u\), up to a prescribed residual factor \(h\). The
large increasing factor \(e^{G(x)}\) and the large decreasing factor
\(e^{-G(x)}\) cancel, leaving exactly the perturbation \(h(x)\). This
shows that the biphasic behavior in the scaffold system is not a formal
consequence of multiplying increasing and decreasing quantities. Rather,
it is a consequence of the particular steady-state equations, which
force the logarithmic derivative of the fully bound complex to change
sign exactly once.

The case \(m=1\) already illustrates this point within the present class
of models. There is then one ligand, and the bound complex is
\[
y_{\{1\}}^{\maybestar}=K_1y_\varnothing^{\maybestar}a_1^{\maybestar}.
\]
As \(X=Y_{\mathrm{tot}}\) increases, the free scaffold
\(y_\varnothing^{\maybestar}\) increases strictly, while the free ligand
\(a_1^{\maybestar}\) decreases strictly. Nevertheless, the product
\(K_1y_\varnothing^{\maybestar}a_1^{\maybestar}\) is not biphasic: by
Theorem~\ref{theo:monotonefree}, it is strictly increasing. In fact,
\[
y_{\{1\}}^{\maybestar}(X)\to A_{1,\mathrm{tot}}
\qquad\text{as }X\to\infty.
\]
Thus the product of the increasing free scaffold concentration and the
decreasing free ligand concentration need not have an interior maximum.

\end{remark}

\section{Discussion and further remarks}
\label{sec:remarks}

There are several possible directions for future work, including the study of
the transient dynamics of the system under time-varying, or even simply
periodic, scaffold availability. We collect below some additional remarks.

\begin{remark}
In the special case $m=2$, the steady-state equations
\eqref{eq:Ytot_product_general}--\eqref{eq:Aitot_product_general}
can be solved explicitly. Write
\[
X:=Y_{\mathrm{tot}},
\qquad
\alpha_i:=K_iA_{i,\mathrm{tot}},
\qquad i=1,2,
\]
and define
\[
P_i
:=
\sqrt{(K_iX-\alpha_i-1)^2+4K_iX},
\qquad i=1,2.
\]
Equivalently,
\[
P_i
=
\sqrt{(\alpha_i+1-K_iX)^2+4K_iX}.
\]
Then the positive steady state is given by
\[
a_i^{\maybestar}
=
\frac{\alpha_i-K_iX-1+P_i}{2K_i},
\qquad i=1,2,
\]
and
\[
y_\varnothing^{\maybestar}
=
\frac{(\alpha_1+1-K_1X-P_1)(\alpha_2+1-K_2X-P_2)}
{4K_1K_2X}.
\]
These formulas are obtained by solving the quadratic equations
\eqref{eq:Ui_quadratic_general} for
\[
U_i=1+K_ia_i,
\]
namely
\[
U_i^{\maybestar}
=
\frac{\alpha_i+1-K_iX+P_i}{2},
\qquad i=1,2,
\]
and then using
\[
a_i^{\maybestar}=\frac{U_i^{\maybestar}-1}{K_i},
\qquad
y_\varnothing^{\maybestar}=\frac{X}{U_1^{\maybestar}U_2^{\maybestar}}.
\]
The displayed formula for $y_\varnothing^{\maybestar}$ follows by rationalizing the
denominators.

The steady-state concentration of the fully bound complex is
\[
y_{\{1,2\}}^{\maybestar}
=
y_\varnothing^{\maybestar}K_1K_2a_1^{\maybestar}a_2^{\maybestar}.
\]
Using the preceding expressions, or equivalently the identity
\[
\frac{U_i^{\maybestar}-1}{U_i^{\maybestar}}
=
\frac{K_iX+\alpha_i+1-P_i}{2K_iX},
\]
one obtains
\[
y_{\{1,2\}}^{\maybestar}
=
\frac{(K_1X+\alpha_1+1-P_1)(K_2X+\alpha_2+1-P_2)}
{4K_1K_2X}.
\]
When $K_1=K_2=1$, this reduces to
\[
y_{\{1,2\}}^{\maybestar}
=
\frac{(X+A_{1,\mathrm{tot}}+1-P_1)
      (X+A_{2,\mathrm{tot}}+1-P_2)}
{4X},
\]
with
\[
P_i=\sqrt{(X-A_{i,\mathrm{tot}}-1)^2+4X}.
\]
This is the same expression as the corresponding formula in
\cite{Douglass2013ThreeBody}. In the notation of that reference, the
formula for the free scaffold concentration also agrees with their
equation~(S57), after simplification, except that the denominator should
contain $X$ rather than $X^2$.
\end{remark}

\begin{remark}
The independent binding assumption may not always be
valid in applications.  For example, in the synthetic biology example
illustrated in Figure~\ref{fig:trispecific}, promoter binding might be
facilitated by the previous binding of the two other components.
Also, depending on the time scale of the processes,
one might want to account for production, degradation, and dilution of
the various players (ligands and/or scaffold). These additional
processes are particularly relevant in that example, and are included
in the model studied in~\cite{ManojDelVecchio2022CDC,24cdc_activator_repressor,26_activator_repressor_journal}.  It would be
interesting to extend the theorems presented here to such cases, including
allosteric or context-dependent binding.

\end{remark}

\subsection*{Acknowledgements}
This research was supported in part by AFOSR grants
FA9550-21-1-0289 and FA9550-22-1-0316.

\siamls{%
  \newpage

\subsubsection*{Declarations (AI content)}

During the preparation of this manuscript, AI-assisted technologies
(various versions of ChatGPT, Gemini, Claude, and Grok utilized
between mid-2025 and mid-2026) were employed to verify algebraic
computations, review logical arguments, and refine the writing. MATLAB
and Python scripts were also utilized for computational verification
and numerical exploration. The author assumes full responsibility for
the accuracy and integrity of all content. 

\subsection*{Appendix: Relevant mathematical background}

\subsubsection*{Biological interaction networks}

Biological interaction networks provide a mathematical framework for
modeling the dynamics of interacting species.  The species may represent,
for example, molecules such as mRNA, proteins, or metabolites in an
intracellular network, or classes of individuals such as susceptible,
infected, and recovered individuals in an epidemiological model, or
predators and prey in an ecological model.  We review here some standard
terminology; see, for example,~\cite{MAli_angeli_sontag_contractions_learn}
for a recent reference.

Formally, a biological interaction network (BIN), or chemical reaction
network (CRN), is specified by a set of \emph{species}
\[
X_1,\ldots,X_n
\]
and a set of \emph{reactions}
\[
R:=\{\bR_1,\ldots,\bR_\nu\}.
\]
Each reaction is written symbolically as
\[
\bR_j:
\sum_{i=1}^n \alpha_{ij}X_i
\longrightarrow
\sum_{i=1}^n \beta_{ij}X_i,
\]
where the coefficients \(\alpha_{ij},\beta_{ij}\) are nonnegative
numbers called \emph{stoichiometric coefficients}.  The net gain, or
loss, of the \(i\)th species in the \(j\)th reaction is
\[
\gamma_{ij}:=\beta_{ij}-\alpha_{ij}.
\]
These numbers are collected into the stoichiometry matrix
\(\Gamma\in\mathbb R^{n\times \nu}\), defined entrywise by
\[
[\Gamma]_{ij}=\gamma_{ij}.
\]

To quantify the state of the network, the species
\(X_1,\ldots,X_n\) are assigned nonnegative numbers
\(x_1,\ldots,x_n\), called their \emph{concentrations}.  The reactions
\(\bR_1,\ldots,\bR_\nu\) are assigned \emph{rates}
\[
R_j:\mathbb R_{\ge0}^n\to \mathbb R_{\ge0},
\qquad j=1,\ldots,\nu,
\]
which are assumed to be locally Lipschitz.  The most common choice of
reaction rates is \emph{mass-action kinetics}, in which
\[
R_j(x)=k_j\prod_{i:\,\alpha_{ij}>0}x_i^{\alpha_{ij}},
\]
where \(k_j>0\) is the kinetic constant of the \(j\)th reaction.

The time evolution of the concentration vector
\[
x=[x_1,\ldots,x_n]^T
\]
from an initial concentration vector \(x_0\) is described by the ordinary
differential equation
\[
\dot x=\Gamma R(x),\qquad x(0)=x_0.
\]
This ODE is a \emph{positive system}: if the initial condition is
nonnegative, then the trajectory remains nonnegative for all future time.

A nonzero vector \(w\in\mathbb R_{\ge0}^n\) is called a
\emph{conservation law} if
\[
w^T\Gamma=0.
\]
It corresponds to a conserved quantity, since along solutions
\[
w^Tx(t)\equiv w^Tx(0).
\]
Let \(G\) denote the image, or column space, of \(\Gamma\).  Solutions
are confined to translates of \(G\).  Indeed, integrating the ODE gives
\[
x(t)=x_0+\Gamma\int_0^t R(x(\tau))\,d\tau,
\]
and therefore
\[
x(t)\in S_{x_0}
\qquad \text{for all } t\ge0,
\]
where
\[
S_{x_0}:=(\{x_0\}+G)\cap\mathbb R_{\ge0}^n.
\]
The invariant set \(S_{x_0}\) is called the \emph{stoichiometric
compatibility class} corresponding to the initial condition \(x_0\).

}

\newpage
\bibliographystyle{abbrvnat}
\bibliography{scaffolds}

@article{Wu2020_Trispecific_NatCancer,
  author       = {Lan Wu and Edward Seung and Ling Xu and Ercole Rao and Dana M. Lord and Ronnie R. Wei and Virna Cortez‑Retamozo and Beatriz Ospina and Valeriya Posternak and Gregory Ulinski and Peter Piepenhagen and Elisa Francesconi and Nizar El‑Murr and Christian Beil and Patrick Kirby and Aiqun Li and Jennifer Fretland and Rita Vicente and Gejing Deng and Tarik Dabdoubi and Beatrice Cameron and Thomas Bertrand and Paul Ferrari and Stéphanie Pouzieux and Cendrine Lemoine and Catherine Prades and Anna Park and Huawei Qiu and Zhili Song and Bailin Zhang and Fangxian Sun and Marielle Chiron and Srinivas Rao and Katarina Radošević and Zhi‑yong Yang and Gary J. Nabel},
  title        = {Trispecific antibodies enhance the therapeutic efficacy of tumor‑directed {T} cells through {T} cell receptor co‑stimulation},
  journal      = {Nature Cancer},
  volume       = {1},
  pages        = {86--98},
  year         = {2020},
  doi          = {10.1038/s43018-019-0004-z}
}

@book{Feinberg2019FCRNT,
  author    = {Martin Feinberg},
  title     = {Foundations of Chemical Reaction Network Theory},
  year      = {2019},
  series    = {Applied Mathematical Sciences},
  volume    = {202},
  publisher = {Springer},
  address   = {Cham},
}

@article{persistencePetri,
 year = 2007,
 author = {D. Angeli and P. de Leenheer and E. D. Sontag},
 title = {A {P}etri net approach to the study of persistence in chemical reaction networks},
 journal= {Mathematical Biosciences},
 volume={210},
  pages={598-618},
}

@article{Tcell01,
    AUTHOR = {Sontag, E. D.},
 TITLE = {Structure and stability of certain chemical networks and
              applications to the kinetic proofreading model of {T}-cell
              receptor signal transduction},
   JOURNAL = {IEEE Trans. Automat. Control},
    VOLUME = {46},
      YEAR = {2001},
    NUMBER = {7},
     PAGES = {1028--1047},
      ISSN = {0018-9286},
}

@article{Anderson2008GAS,
  author  = {David F. Anderson},
  title   = {Global Asymptotic Stability for a Class of Nonlinear Chemical Equations},
  journal = {SIAM Journal on Applied Mathematics},
  year    = {2008},
  volume  = {68},
  number  = {5},
  pages   = {1464--1476},
  publisher = {Society for Industrial and Applied Mathematics}
}

@article{2024_bispecific,
author = {M. Sadeghi and I. Kareva and G. Pogudin and E. D. Sontag},
title = {Quantitative pharmacology methods for bispecific {T} cell engagers},
year = {2025},
journal = {Bulletin of Mathematical Biology},
volume = {87},
pages = {85},
}

@article{SiegelMacLean2000,
  author       = {Siegel, D. and MacLean, D.},
  title        = {Global stability of complex balanced mechanisms},
  journal      = {Journal of Mathematical Chemistry},
  volume       = {27},
  number       = {1--2},
  pages        = {89--110},
  year         = {2000},
  month        = {October},
  doi          = {10.1023/A:1019183206064},
}

@inproceedings{ManojDelVecchio2022CDC,
  author    = {Krishna Manoj and Domitilla Del Vecchio},
  title     = {Emergent interactions due to resource competition in {CRISPR}-mediated genetic activation circuits},
  booktitle = {2022 IEEE 61st Conference on Decision and Control (CDC)},
  pages     = {1300--1305},
  year      = {2022},
  publisher = {IEEE},
}

@article{Douglass2013ThreeBody,
  author       = {Eugene F. {Douglass Jr} and Chad J. Miller and Gerson Sparer and Harold Shapiro and David A. Spiegel},
  title        = {A comprehensive mathematical model for three-body binding equilibria},
  journal      = {Journal of the American Chemical Society},
  year         = {2013},
  volume       = {135},
  number       = {16},
  pages        = {6092--6099},
  doi          = {10.1021/ja311795d}
}

@article{Yang2011ScaffoldmediatedNO,
  title={Scaffold-mediated Nucleation of Protein Signaling Complexes: Elementary Principles},
  author={J. Yang and W.S. Hlavacek},
  journal={Mathematical Biosciences},
  year={2011},
  volume={232},
  pages={164-173}
}

@INPROCEEDINGS{24cdc_activator_repressor,
  author = {M. Ali Al-Radhawi and K. Manoj and D. Jatkar and A. Duvall and D. {Del Vecchio} and E.D. Sontag},
  title = {Competition for binding targets results in paradoxical effects for simultaneous activator and repressor action},
  booktitle = {Proc. 63rd IEEE Conference on Decision and Control (CDC)},
  year = {2024},
  pages = {5579-5585},
  }

@article{26_activator_repressor_journal,
author = {K. Manoj and D. Jatkar and M. Ali Al-Radhawi and E.D. Sontag and D. {Del Vecchio}},
title = {Paradoxical gene regulation explained by competition for genomic sites},
journal = {},
year = {2026},
note = {Submitted. Preprint available at bioRxiv 10.1101/2025.11.27.691022.},
  }

@misc{genscript,
note = {https://www.genscript.com/bispecific-antibody.html},
}

@ARTICLE{dunlapcao,
AUTHOR={Dunlap, Tyler  and Cao, Yanguang },
TITLE={Physiological Considerations for Modeling in vivo Antibody-Target Interactions},
JOURNAL={Frontiers in Pharmacology},
VOLUME={Volume 13 - 2022},
YEAR={2022},
DOI={10.3389/fphar.2022.856961},
ISSN={1663-9812}
}

@article{GoodZalatanLim2011,
  title = {Scaffold proteins: hubs for controlling the flow of cellular information},
  author = {Good, Matthew C. and Zalatan, Jesse G. and Lim, Wendell A.},
  journal = {Science},
  volume = {332},
  number = {6030},
  pages = {680--686},
  year = {2011},
  month = {May},
  doi = {10.1126/science.1198701},
  pmid = {21551057},
  pmcid = {PMC3117218},
  publisher = {American Association for the Advancement of Science}
}

@article{ShawFilbert2009,
  title = {Scaffold proteins and immune-cell signalling},
  author = {Shaw, Andrei and Filbert, Elizabeth},
  journal = {Nature Reviews Immunology},
  volume = {9},
  number = {1},
  pages = {47--56},
  year = {2009},
  month = {January},
  doi = {10.1038/nri2473},
  pmid = {19104498},
  publisher = {Nature Publishing Group}
}

\end{document}